\documentclass[reprint, aps, twocolumn]{revtex4-2}
\usepackage{natbib}

\input epsf
\usepackage{graphicx}
\usepackage{dcolumn}
\usepackage{bm}
\usepackage{amssymb}
\usepackage{amsmath}
\usepackage[super]{nth}
\usepackage{multirow}
\usepackage{array}

\usepackage{subcaption}
\usepackage{amsfonts,yfonts,mathtools}
\usepackage[usenames,dvipsnames,svgnames,table]{xcolor}
\usepackage{gensymb}

\newcommand{\nn}{\noindent}

\newcommand{\ra}{\rightarrow}

\newcommand{\be}{\begin{equation}}
\newcommand{\ee}{\end{equation}}
\newcommand{\been}{\begin{enumerate}}
\newcommand{\enen}{\end{enumerate}}
\newcommand{\beit}{\begin{itemize}}
\newcommand{\enit}{\end{itemize}}

\newcommand{\mbf}{\mathbf}
\newcommand{\mrm}{\mathrm}
\newcommand{\mcl}{\mathcal}

\DeclareMathOperator{\Tr}{Tr}

\begin{document}

\title{Entangling Solid Solutions: Machine Learning of Tensor Networks for Materials Property Prediction}
\author{David~E.~Sommer}
\affiliation{Department of Electrical Engineering, University of Washington, Seattle, WA, 98195, USA}
\author{Scott~T.~Dunham}
\affiliation{Department of Electrical Engineering, University of Washington, Seattle, WA, 98195, USA}

%----------------------------------------------------------------------------------------
\begin{abstract}
Progress in the application of machine learning techniques to the prediction of solid-state and molecular materials properties has been greatly facilitated by the development state-of-the-art feature representations and novel deep learning architectures.  A large class of atomic structure representations based on expansions of smoothed atomic densities have been shown to correspond to specific choices of basis sets in an abstract many-body Hilbert space.  Concurrently, tensor network structures, conventionally the purview of quantum many-body physics and quantum information, have been successfully applied in supervised and unsupervised learning tasks in computer vision and natural language processing.  In this work, we argue that architectures based on tensor networks are well-suited to machine learning on Hilbert-space representations of atomic structures.  This is demonstrated on supervised learning tasks involving widely available datasets of density functional theory calculations of metal and semiconductor alloys.  In particular, we show that certain standard tensor network topologies exhibit strong generalizability even on small training datasets while being parametrically efficient.  We further relate this generalizability to the presence of complex entanglement in the trained tensor networks.  We also discuss connections to learning with generalized structural kernels and related strategies for compressing large input feature spaces.
\end{abstract}
%----------------------------------------------------------------------------------------

\maketitle

%----------------------------------------------------------------------------------------
%----------------------------------------------------------------------------------------
\section{Introduction} \label{intro}
Rapid advances in electronic structure methods and computational resources have enabled high-throughput \textit{ab initio} calculations of solids and molecules over broad classes of chemistries and structures.  Much of this work has been motivated by a pressing need to expand the tools for engineering new materials with desirable properties, with applications ranging from drug design to optoelectronics, energy storage, and quantum computing.  First-principles calculations, in particular, play a crucial role in mapping the atomistic structure and composition of a material to fundamental properties such as ground state energies, potential energy surfaces, band structures and optical excitation spectra.  However, a full exploration of materials design space is plagued by various well-known curses of dimensionality.  These include the exponential scaling of many-body Hilbert spaces and the associated computational complexity of solving the many-body Schr\"{o}dinger equation, as well as the combinatorics of decorating finite and periodic lattices by various chemical species.  While density functional theory (DFT) and many-body perturbation theory (MBPT) have seen numerous successes in addressing the former, they can still present large computational barriers to address the latter.

In recognition of this challenge, the development and application of machine learning techniques to produce accurate and computationally efficient surrogate models of \textit{ab initio} calculations has become a very active area of research \cite{Behler:prl.2007, Bartok:prb.2013, Thompson:jcp.2015, Shapeev:mms.2016, Huo:arxiv.2017, Shapeev:cms.2017, Drautz:prb.2019, Willatt:jcp.2019, Behler:chemrev.2021, Nyshadham:npj-cm.2019, Sutton:npj-cm.2019, Langer:arxiv.2020}.  Broadly speaking, the success of these techniques depends on the identification of a sufficiently descriptive feature space which captures the variance of the data one wishes to model.  Much progress has been made recently in identifying and engineering feature spaces which accurately represent local atomic environments and can be used as inputs to standard machine learning algorithms.  Efficient representations are often achieved by constructions that respect physical symmetries, such as invariance to global rotations, translations and permutations of identical chemical species.  For the class of atom-centered representations based on expansions of atomic densities, symmetry is incorporated either by directly constructing invariant polynomials, as is the case for atom-centered symmetry functions (ACSFs) \cite{Behler:prl.2007} and moment tensor potentials (MTPs) \cite{Shapeev:mms.2016}, or by explicit integration over relevant symmetry groups, as exemplified by the smooth overlap of atomic positions (SOAP) \cite{Bartok:prb.2013} and spectral neighbor analysis potentials (SNAP) \cite{Thompson:jcp.2015}.  To a large extent, instances in this class of atom-centered representations correspond to different choices of basis sets in an abstract $N$-body Hilbert space \cite{Willatt:jcp.2019}, and the hierarchy of $N$-body features has recently been shown to be organized in the so-called atomic cluster expansion (ACE) framework \cite{Drautz:prb.2019, Drautz:prb.2020, Bachmayr:arxiv:2019, Lysogorskiy:npj_cm.2021}.

It is worth mentioning that an alternative to extensive feature engineering is to employ a so-called \textit{end-to-end} approach, in which inputs to the model are minimally processed, and the relevant feature space is learned by the model architecture during the course of training.  Examples of this approach in materials informatics include graph neural network methods  \cite{Schutt:JCP.2018, Xie:PRL.2018, Chen:ChemMat.2019, Klicpera:arxiv2020} and explicitly symmetry-equivariant architectures \cite{Thomas:arxiv.2018, Fuchs:ANIPS.2020}.  However, in the absence of prior feature engineering, training accurate deep learning models becomes difficult when the amount of data is limited.

Thus, from a practical standpoint, balancing the bias and complexity of a model between the two extremes of extensive feature engineering and highly flexible model architectures is often influenced by the availability of training data.  Strictly end-to-end deep learning can require substantially more data when correlations between features are complex, while naive application of parametrically efficient features can lead to a model which generalizes poorly on new inputs.  Of course, the strict distinction between these two extremes can be somewhat arbitrary: the latent space encoded by the hidden layers of a deep learning architecture can be thought of as a kind of renormalized feature space \cite{Mehta:arxiv.2014, Lin:jsp.2016}, subject to its own forms of bias through choices of hyperparameters, regularization schemes and architectural topology.  To that end, certain deep model architectures may possess better so-called \textit{inductive bias} \cite{Battaglia:arxiv.2018} over a given dataset, more efficiently prioritizing the search space of solutions when their structure reflects the pattern of correlations in the feature space.

Interesting connections between inductive bias and feature correlations have grown out of recent efforts to introduce techniques from the study of quantum entanglement and strongly correlated quantum many-body systems to the field of machine learning.  Specifically, tensor network methods have been successfully applied to supervised and unsupervised learning tasks in computer vision and natural language processing \cite{Novikov:arxiv.2015, Stoudenmire:arxiv.2016, Han:PRX.2018, Efthymiou:arxiv.2019, Martyn:arxiv.2020, Glasser:IEEEa.2020, Wang:arxiv.2020, Gao:prr.2020, Reyes:arxiv.2020, Cheng:PRB.2021} and have provided a basis for analyzing the expressiveness of common deep learning architectures based on their entanglement properties \cite{Levine:arxiv.2017, Levine:prl.2019}.  Indeed, a general argument for the effectiveness of tensor networks in machine learning contexts is that the pattern of entanglement encoded in the network can efficiently represent the pattern of correlation in the feature space of the data.

In this work, we argue for the applicability of tensor networks in machine learning structure-property models of materials.  This problem is addressed from two directions: we show how both the construction of an input feature space and of an associated machine learning architecture can be formulated in the language of tensor networks.  In Section \ref{so3-reps}, a large class of atomic structure representations, corresponding to the SO(3)-invariant tensor basis set of the (smoothed) atomic cluster expansion (ACE), are shown to admit a natural tensor network description.  An immediate consequence of this rewriting is that the equivalence classes and the recursive construction of the hierarchy of $N$-body ACE basis tensors become transparent in the graphical notation of the tensor network.  In Section \ref{tensor-facts}, we show how common tensor network factorizations for the weights of machine learning models can be naturally built on top of individual ACE basis tensors.  We discuss the relationship between this approach and kernel learning, and in particular how certain tensor networks can realize a form of \textit{alchemical learning} by coupling information between the local atomic structure and the chemical elements within it.  We also comment on how the introduction of copying and merging operations in tensor network structures can be used to introduce higher-order correlations from fixed $N$-body features, akin to the construction of nonlinear kernels.  

In Section \ref{tn-benchmark}, we benchmark some of the proposals from Sections \ref{so3-reps} and \ref{tensor-facts} using widely available datasets of density functional theory calculations of metal and semiconductor alloys.  For concreteness, we use input features corresponding to the commonly used SOAP power spectrum and study the learning performance of different tensor network factorizations of the model weights on training sets of various sizes.  Compared to standard kernel learning methods and fully-connected neural networks, we find that models based on matrix product states (MPS) and matrix product operators (MPO) show strong generalizability and parametric efficiency, with notable performance on small training sets.  We subsequently study the entanglement properties of the trained networks and find signatures of high entanglement complexity consistent with the models' strong generalizability.  We also provide some evidence that the latent spaces learned by the hidden layers of the networks are able to capture physically relevant structural and chemical information, and we utilize this insight to effectively compress the input basis tensors.

%----------------------------------------------------------------------------------------
%----------------------------------------------------------------------------------------
\section{SO(3)-Invariant Atomic Representations} \label{so3-reps}
In this work, we are concerned with the prediction of some atomic property $V$ given a configuration $\{\mbf{r}_i\}$ of N atomic species.  For simplicity, we will restrict to the case where $V$ is a scalar, such as the total energy or band gap of the system, but extensions to vectorial and tensorial properties are discussed in Appendix \ref{appendix:equivariants}.  Following the discussions by Drautz \cite{Drautz:prb.2019} and Lysogorskiy \textit{et al} \cite{Lysogorskiy:npj_cm.2021}, a natural starting point is to formulate a coarse-grained model given by a sum of local terms, $V = V^{(0)} + \sum_i V_i$, where the local property $V_i$ associated with atom $i$ can be approximated by expanding in terms of local $n$-body interactions $V^{(n)}$:

	\begin{align}
		V_i &= V^{(1)}(\mbf{r}_i) +\frac{1}{2}\sum_j V^{(2)}(\mbf{r}_i, \mbf{r}_j) \nonumber \\
		&+ \frac{1}{3!}\sum_{jk} V^{(3)}(\mbf{r}_i, \mbf{r}_j, \mbf{r}_k) + \dotsb
	\label{local_prop}
	\end{align}

\nn
Requiring consistency with fundamental symmetries introduces constraints on the form of the interactions.  For instance, translation invariance leads to atom-centered functions of the form $V^{(\nu+1)}(\mbf{r}_{j_1 i},\dotsc,\mbf{r}_{j_\nu i})$, where $\mbf{r}_{ji} = \mbf{r}_{j} - \mbf{r}_{i}$ can be thought of as an effective bond from the central site $i$ to a site $j$ in its relative environment.  Interaction terms can be further decomposed by projecting onto an appropriate $\nu$-order basis set $\Phi_{s_1,s_2,\dotsc,s_\nu}$,

	\begin{align}
		V^{(\nu+1)}(\mbf{r}_{j_1 i},&\dotsc,\mbf{r}_{j_\nu i}) = \nonumber \\
		& \sum_{\{s\}} J_{s_1,s_2,\dotsc,s_\nu} \Phi_{s_1,s_2,\dotsc,s_\nu}(\mbf{r}_{j_1 i},\dotsc,\mbf{r}_{j_\nu i}) \ ,
	\label{mb_interaction}
	\end{align}

\nn
with general interaction coefficients $J_{s_1,s_2,\dotsc,s_\nu}$.  A crucial insight from the recent development of atom-centered descriptors \cite{Bartok:prb.2013, Drautz:prb.2019, Willatt:jcp.2019, Bachmayr:arxiv:2019} is that an efficient representation of atomic environments can be obtained by a low-rank approximation of the $\Phi_{s_1,s_2,\dotsc,s_\nu}$ cluster basis in terms of a single-bond basis $\phi_{s_k}(\mbf{r}_{j_k i})$,

	\be
		\Phi_{s_1,s_2,\dotsc,s_\nu}(\mbf{r}_{j_1 i},\dotsc,\mbf{r}_{j_\nu i}) = \prod_{k=1}^{\nu} \phi_{s_k}(\mbf{r}_{j_k i}) \ ,
	\label{mb_bond_basis}
	\ee
		
\nn
combined with a reorganization of the summations in (\ref{local_prop}) and (\ref{mb_interaction}),

	\be
		\sum_{j_1 \dotsm j_\nu} \Phi_{s_1,s_2,\dotsc,s_\nu}(\mbf{r}_{j_1 i},\dotsc,\mbf{r}_{j_\nu i}) = \prod_{k=1}^{\nu} A_{i, s_k} \ .
	\label{mb_atomic_basis}
	\ee
	
\nn
We note that the unrestricted summation in (\ref{mb_atomic_basis}) contains products over identical bonds.  In the ACE framework \cite{Drautz:prb.2019}, these self-interaction terms can be canceled by lower order terms in the expansion (\ref{local_prop}).

Formally speaking, the atom-centered descriptor, $A_{i, s} \coloneqq \sum_{j} \phi_{s}(\mbf{r}_{j i})$, can be understood as the expansion coefficients of the local atomic density $\sigma_i$ in an abstract atom-centered Hilbert space $| s \rangle \in \mcl{V}$,

	\be
		| \sigma_i \rangle = \sum_{s} A_{i,s} | s \rangle \ .
	\label{ac_basis}
	\ee

\nn
In practice, the real-space atomic density for a given atomic environment is approximated by a superposition of localized functions $h$,

	\be
		\langle \mbf{\alpha r} |\sigma_i \rangle  = \sum_{j \neq i} \delta_{\alpha \alpha_j} f_c(r_{ji}) h(\mbf{r} - \mbf{r}_{ji}) \ ,
	\label{smooth_density}
	\ee

\nn
where species $\alpha_j$ occupies site $j$ and $f_c(r_{ji})$ is a smooth cutoff function that restricts the sum to some local environment of atom $i$.  In the original ACE formalism \cite{Drautz:prb.2019}, $h$ is chosen to be a delta function, whereas in the SOAP and SNAP formalisms \cite{Bartok:prb.2013}, $h$ is chosen to be a Gaussian.  

The atom-centered basis (\ref{ac_basis}) is not generally invariant under action of the rotation group $\mcl{G}=$ SO(3).  Rather, according to Maschke's theorem, the atom-centered Hilbert space $\mcl{V}$ decomposes into a direct sum of irreducible representations (``irreps") of SO(3), $\mcl{V} \cong \bigoplus_l \mcl{D}_l \otimes \mcl{V}_l$, where $\mcl{V}_l$ is the subspace of the irrep $l = 0, 1, 2, \dotsc$, and the degeneracy space $\mcl{D}_l$ contains additional degrees of freedom (e.g., atomic species $\alpha$ and purely radial components $n$) which are untouched by the group action.  The generic indices, $s$, of the single-bond basis functions can thus be replaced by the set $(\alpha n l m)$, where $-l \leq m \leq l$ labels components of the irrep subspace $\mcl{V}_l$, and the angular momentum channels of the real-space single-bond basis correspond to spherical harmonics $Y_{l}^{m}(\hat{\mathbf{r}})$,

	\be
		\langle \mbf{r} | \alpha nlm \rangle = \phi_{\alpha nlm}(\mbf{r}) | \alpha \rangle =  R_{n l}(r) Y_{l}^{m}(\hat{\mathbf{r}}) | \alpha \rangle \ .
	\ee
	
\nn
Common choices for radial basis functions $R_{n l}(r)$ are spherical Gaussian-type orbitals and orthogonal polynomials \cite{Bartok:prb.2013, Drautz:prb.2019}.  

The expansion of the local atomic density centered on site $i$ can thus be written as

	\be
		| \sigma_i \rangle = \sum_{\alpha nlm} A_{i,\alpha nlm} | \alpha n l m \rangle \ ,
	\label{irrep_decomp}
	\ee
\nn
where the atom-centered descriptors are given by

	\begin{align}
		A_{i,\alpha nlm} &= \langle \alpha n l m | \sigma_i \rangle \\
		&= \int_{\mathbb{R}^{3}} d\Omega \ R_{n l}(r) Y_{l}^{m}(\hat{\mathbf{r}}) \langle \alpha \mathbf{r} | \sigma_i \rangle \ .
	\label{irrep_descriptors}
	\end{align}

\nn
Furthermore, the reorganized ($\nu+1$)-body expansion (\ref{mb_atomic_basis}) is equivalent to a product state formed by repeated copies of the density $| \sigma_i^{\otimes \nu} \rangle \coloneqq \bigotimes^\nu | \sigma_i \rangle$,
	
	\be
		 \prod_{k=1}^{\nu} A_{i, \alpha_k n_k l_k m_k} = \langle \alpha_1 n_1 l_1 m_1 \dotsm \alpha_\nu n_\nu l_\nu m_\nu | \sigma_i^{\otimes \nu} \rangle \ .
	\label{ac_product_state}
	\ee

\nn
The expressions (\ref{irrep_decomp}) and (\ref{ac_product_state}) suggest that each term in the $\nu$-order series expansion admits a natural description in terms of a tensor network (TN) for the states $| \sigma_i^{\otimes \nu} \rangle$.  In the following, we will consider an associated graphical calculus which incorporates the SO(3) representations carried by the tensors, $A_{i, \alpha_k n_k l_k m_k}$, which will be useful in explicitly constructing SO(3)-invariant descriptors.  In particular, the formalism presented below closely follows recent treatments of symmetric tensor networks \cite{Singh:pra.2010, Singh:prb.2012, Schmoll:ap.2020, Biamonte:aipa.2011} and the earlier, related development of spin networks \cite{Penrose:spinnets.1971, Oeckl:dgt} from angular momentum recoupling theory \footnote{Most of the graphical machinery discussed below carries over more generally to other symmetric tensor categories, of which finite-dimensional representations of SO(3) are an example.}.
	
%----------------------------------------------------------------------------------------
\subsection{Graphical calculus for atomic descriptors} \label{descriptor-calc}
\begin{figure}[!h]
	\centering
		\includegraphics[scale=0.3]{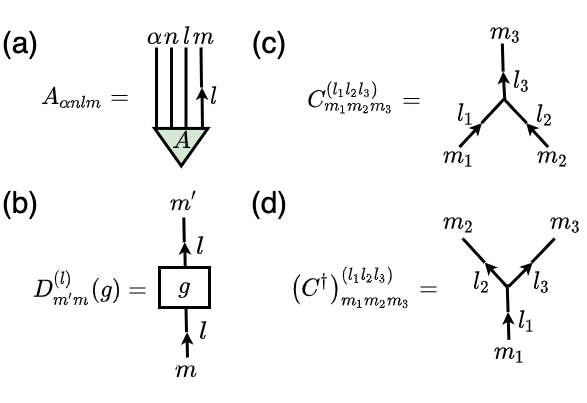}
		\caption{Basic graphical components for the construction of invariant descriptors, including (a) atom-centered descriptors (\ref{irrep_descriptors}), (b) Wigner $D$-matrices, and (c,d) Clebsch-Gordan coefficients as fusion and splitting nodes.}
	\label{basis_network_comps}
	\end{figure}

Working with explicit expressions for tensor product spaces can quickly become cumbersome as the order $\nu$ increases, and this is particularly true when the tensor space possesses a complex internal structure, as is the case for working with SO(3) symmetry.  However, much of the algebraic structure can be encapsulated in a consistent graphical notation, which we utilize in the following.  In the standard diagrammatic notation of tensor networks (Appendix \ref{appendix:tn-basics}), the basic atom-centered descriptor $A_{\alpha n l m}$, as a 4-index tensor, can be represented by a shape with 4 open edges (Figure \ref{basis_network_comps}).  Note that an arrow is added to the edge carrying the irrep space, $\mcl{V}_l$, of the symmetry group to distinguish it from its dual vector space, $\mcl{V}^*_l$, and we will subsequently drop the atom site index $i$ for simplicity.  Regular representations of SO(3) are given by the unitary Wigner $D$-matrices, $D^{(l)}_{m'm}(g)$, and the action of $D^{(l)}_{m'm}(g)$ on the atom-centered basis is given by \textit{contraction} with $A$ along the irrep edge, i.e., by $\sum_{m}D^{(l)}_{m'm}(g) A_{\alpha n l m}$ (cf. Appendix \ref{appendix:tn-basics}).

A $\nu$-order product state of multiple $A$-tensors (\ref{ac_product_state}) is depicted simply by a row of disconnected $A$'s.  This $\nu$-order product state is not invariant under three-dimensional rotations, $U_g^{\otimes \nu} |\sigma_i^{\otimes \nu} \rangle \neq |\sigma_i^{\otimes \nu} \rangle$.  However, it can be made so by explicitly taking the Haar integral over $g \in$ SO(3), $| \sigma_i^{\otimes \nu} \rangle_{g} \coloneqq \int_g U_g^{\otimes \nu} |\sigma_i^{\otimes \nu} \rangle$, since the homomorphism $U_{g_1} U_{g_2} = U_{g_1 g_2}$ extends to the tensor product space. In tensor components, this symmetrized descriptor is given by
		
	\begin{align}
		\langle \alpha_1 n_1 l_1 m_1 &\dotsm \alpha_\nu n_\nu l_\nu m_\nu | \sigma_i^{\otimes \nu} \rangle_{g} \nonumber \\
		& = \int_{g \in\mrm{SO}(3)} d g \prod_{k=1}^{\nu} \langle \alpha_k n_k l_k m_k | U_g | \sigma_i \rangle \ ,
	\label{sym_descriptors}
	\end{align}
	
\begin{figure}[!h]
	\centering
		\includegraphics[scale=0.25]{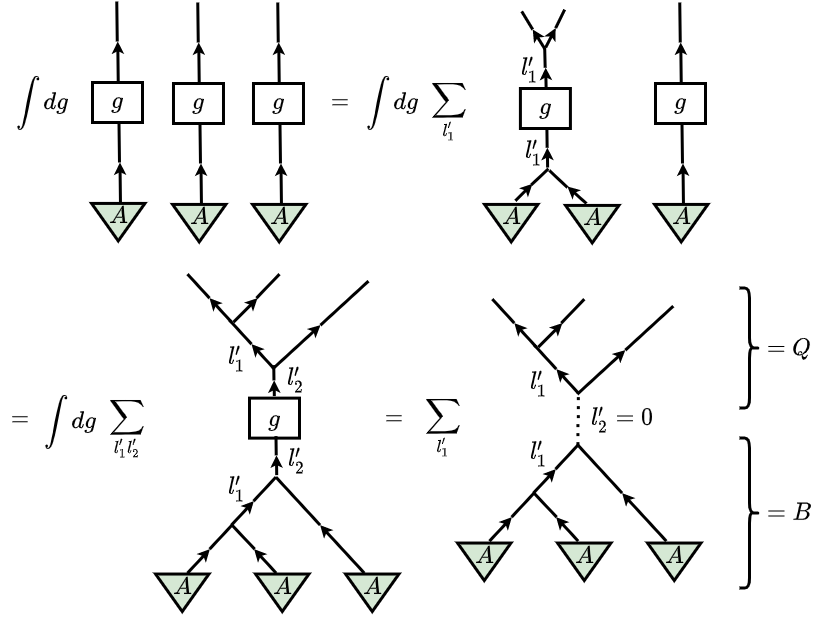}
		\caption{The explicit symmetrization (\ref{sym_descriptors}) of a ($\nu+1$)-body product state (\ref{ac_product_state}) derived from the algebraic rules of SO(3)-recoupling theory.  For simplicity, the edges $(\alpha n l)$ are not shown.}
	\label{JLV_symmetry}
	\end{figure}

\nn
where, upon insertion of resolutions of identity, one obtains a tensor product of Wigner $D$-matrices, $D^{(l)}_{mm'}(g) = \langle l m | U_g | l m' \rangle$, acting on the $A$-tensors.  An explicit formula for (\ref{sym_descriptors}) can be computed by introducing the intertwiners of the symmetry group, the Clebsch-Gordan (CG) coefficients, into the basic building blocks of the graphical calculus (Figure \ref{basis_network_comps}).  We outline this derivation in Figure \ref{JLV_symmetry} for the case $\nu = 3$, but it can be extended by induction to all $\nu$.  In particular, repeated application of the CG identities and their equivariance (Figure \ref{CG_identities}) reduces (\ref{sym_descriptors}) to a single Wigner $D$-matrix contracted with recursively constructed \textit{fusion trees} of CG coefficients.  A single, invariant irrep edge must transform as the trivial representation, hence the Haar integral projects the final $D$-matrix-decorated edge onto the trivial irrep space.  We distinguish an edge carrying the trivial representation ($l=0$) by a dashed line.  Because the trivial representation is one-dimensional ($m=0$) \footnote{More generally, the trivial representation is the unital object in category $\mrm{Rep}(\mcl{G})$ of representations of the group $\mcl{G}$.}, symmetrization yields two disconnected diagrams, which we label $Q$ and $B$ (see below).

\begin{figure}[!h]
	\centering
		\includegraphics[scale=0.3]{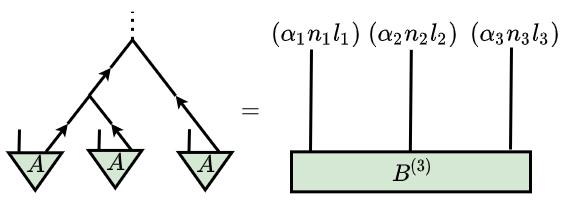}
		\caption{Simplified representation of an invariant descriptor with fixed SO(3)-recoupling scheme.}
	\label{bispect_free_edges}
	\end{figure}

A few comments are in order.  First, the pattern of contractions encapsulated by the structure of the fusion trees in Figure \ref{JLV_symmetry} represents an implicit choice of recoupling scheme.  Different choices of recoupling scheme are possible and are related to each other by unitary transformations, i.e., the so-called $F$-symbols (cf. Appendix \ref{appendix:recoupling}).  Second, symmetrization of $\nu$-order product states for $\nu>2$ introduces summations over intermediate angular momenta $l'$, which we represent explicitly, while contractions of irrep edges imply summation over the corresponding magnetic numbers $m'$.  Finally, this direct symmetrization yields the so-called Jucys-Levinson-Vanagas (JLV) theorems \cite{Stedman:dtgt}, also referred to as the generalized Wigner-Eckhart theorem \cite{Cvitanovic:gt, Singh:pra.2010, Singh:prb.2012, Schmoll:ap.2020}.  Accordingly, a symmetric, $n$-index tensor $T$ decomposes into a tensor product, $T=B \otimes Q$, of a structural tensor $Q$ determined completely by the symmetry group and of a degeneracy tensor $B$.  The structural tensor $Q$, given by a fusion tree of CG coefficients, is known in other contexts as a spin network \cite{Penrose:spinnets.1971, Oeckl:dgt}.  In the spirit of the original Wigner-Eckhart theorem, the degeneracy tensor $B$ is equivalent to the ``reduced matrix element" of the decomposition and encapsulates the degrees of freedom not fixed by the symmetry group \cite{Cvitanovic:gt}.  This is shown explicitly in Figure \ref{bispect_free_edges}, where the $\{\alpha_k n_k l_k\}$ edges remain open.

	\begin{figure}[!h]
	\centering
		\includegraphics[scale=0.25]{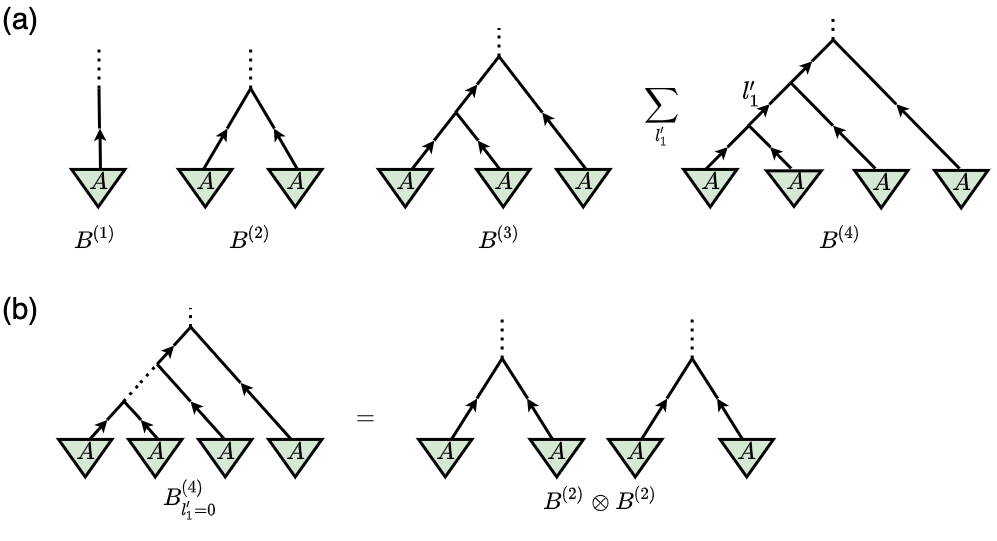}
		\caption{(a) The $\nu$-order hierarchy of SO(3)-invariant descriptors in terms of recursively constructed fusion trees.  (b) Higher order tensors with trivial intermediate irreps factorize into products of lower order tensors.}
	\label{ladder_of_descriptors}
	\end{figure}
	
The $B$-tensors constitute the set of SO(3)-invariant descriptors which form the basis of the ACE framework.  In the recoupling scheme chosen in Figure \ref{JLV_symmetry}, the $\nu$-order series of invariant tensors is given by the contraction of $\nu$ atom-centered tensors $A$ with the appropriate fusion tree (Figure \ref{ladder_of_descriptors}), and invariance is thus manifested by mapping the tensor product of $\nu$ irreps to the trivial representation.  It is readily apparent that invariant descriptors with trivial intermediate irreps factorize, e.g., $B^{(4)}_{l'_1=0} = B^{(2)} \otimes B^{(2)}$.  Moreover, this construction can be generalized \cite{Drautz:prb.2020} to SO(3)-equivariant descriptors $B^{(\nu)}_L$ by allowing the tensor product of $\nu$ irreps to fuse to a non-trivial representation $L$, where the open $L$ edge of the fusion tree carries the dimension of the irrep, $2 L + 1$.  Higher-order SO(3)-equivariant descriptors, $B^{(\nu+1)}_L$, can be built recursively from lower orders $B^{(\nu)}_{L'} \otimes B^{(1)}_{L''}$ by contracting with the so-called \textit{cup} and \textit{cap} tensors (normalized $2jm$ symbols), which diagrammatically enable the orientation of irrep edges to be reversed.  This recursive construction is discussed in more detail in Appendix \ref{appendix:equivariants}.  In the following sections, we will represent the invariant ACE basis tensors $B^{(\nu)}$, as in Figure \ref{bispect_free_edges}, by suppressing the fusion trees.

We give explicit formulae \cite{Drautz:prb.2019, Drautz:prb.2020} for the first few invariant descriptors, while higher order $B^{(\nu)}$ can be constructed via the recursive procedure outlined in Appendix \ref{appendix:equivariants} or simply read off of the corresponding fusion tree:

	\begin{align}
		B^{(1)}_{\alpha n} &= A_{\alpha n 00} \label{trivial_spectrum} \\
		B^{(2)}_{\substack{\alpha_1\alpha_2 \\ n_1 n_2 l}} &= \sum_{m=-l}^{l} \frac{(-1)^{l-m}}{\sqrt{d_l}} A_{\alpha_1 n_1 l m} A_{\alpha_2 n_2 l -m} \label{power_spectrum} \\
		(B^{(3)})_{\substack{n_1 n_2 n_3 \\ l_1 l_2 l_3}}^{\alpha_1 \alpha_2 \alpha_3}  &=  \sum_{m_1 m_2 m_3} 
		     \frac{(-1)^{l_3-m_3}}{\sqrt{d_{l_3}}} C^{(l_1 l_2 l_3)}_{m_1 m_2 -m_3} \nonumber \\
		&\times A_{\alpha_1 n_1 l_1 m_1} A_{\alpha_2 n_2 l_2 m_2} A_{\alpha_3 n_3 l_3 m_3} \label{bispectrum}
	\end{align}
	
\nn
Here, $d_l = 2l+1$ is the dimension of the irrep, and the angular momentum coupling in (\ref{bispectrum}) can be rewritten in terms of the Wigner $3jm$ symbol, as in \cite{Drautz:prb.2019}.  As expected, the $\nu=1$ term loses all angular information after integrating over the SO(3) rotation group.  Hence, accurate expansions of atom-centered properties typically require higher-order terms.  In the $\nu=2,3$ terms, one finds the SOAP power spectrum and bispectrum, respectively \cite{Bartok:prb.2013, Drautz:prb.2019}.  In a real-space basis, these $\nu$-order invariants can be understood as spherical moments of the bond distribution centered on atom $i$.  For $\nu=1$, this corresponds to spherically averaging over a single bond $\mbf{r}_{j_1 i}$, which yields a completely isotropic function depending only on the bond length.  The power spectrum, $\nu=2$, measures the correlation between pairs of bonds ($\mbf{r}_{j_1 i}$, $\mbf{r}_{j_2 i}$), where spherical averaging over the local environment results in a function depending on the bond lengths ($r_{j_1 i}$, $r_{j_2 i}$) and the relative angle, $\hat{\mathbf{r}}_{j_1 i} \cdot \hat{\mathbf{r}}_{j_2 i}$, between bond vectors.  Similarly, a real-space projection of the bispectrum, $\nu=3$, depends only on three bond lengths ($r_{j_1 i}$, $r_{j_2 i}$, $r_{j_3 i}$) and the relative angles between each pair of bond vectors.

%----------------------------------------------------------------------------------------
\section{Tensor Network Learning} \label{tensor-facts}
The SO(3)-invariant, atom-centered descriptors $B_i^{(\nu)}$ can be used as input features for a variety of machine-learning methods.  In the ACE formalism, this amounts to recasting the expansion (\ref{local_prop}) in the symmetrized basis,

	\begin{align}
		V_i &= \sum_\nu w_i^{(\nu)} \cdot B_i^{(\nu)} \nonumber \\
		&= \sum_\nu \sum_{\{ \alpha_k n_k l_k \}} w^{(\nu)}_{i \{ \alpha_k n_k l_k \}} B^{(\nu)}_{i \{ \alpha_k n_k l_k \}} \ ,
	\label{ACE}
	\end{align}

\nn
where the model weights $w_i^{(\nu)}$ are fully contracted with the free edges, $\{\alpha_k n_k l_k\}$, of the invariant descriptors.  As dense tensors, the size of the weights $w_i^{(\nu)}$ scale as $\mcl{O}(|\alpha|^\nu |n|^\nu |l|^\nu)$ with the number of chemical species $|\alpha|$ and the number of radial $|n|$ and angular momentum $|l|$ channels included in the input descriptors.

	\begin{figure}[!h]
	\centering
		\includegraphics[scale=0.3]{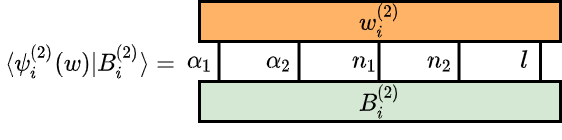}
		\caption{($\nu=2$)-order term in the ACE expansion as an inner product between the descriptor state, $| B_i^{(\nu)} \rangle$, and a learnable state, $|\psi^{(\nu)}_i(w) \rangle$.}
	\label{ace_tensor}
	\end{figure}

It is clear from (\ref{ACE}) that each $\nu$-order term can be understood as an inner product, $\langle \psi^{(\nu)}_i(w) | B_i^{(\nu)} \rangle$, between the invariant basis states $| B_i^{(\nu)} \rangle$ (the reduced part of the symmetrized state $| \sigma_i^{\otimes \nu} \rangle_g$) and a state $|\psi^{(\nu)}_i(w) \rangle$ which depends on a set of learnable parameters $w$ (Figure \ref{ace_tensor}).  We can generalize this construction by considering a tensor network ansatz for $|\psi^{(\nu)}_i(w) \rangle$, where the topology of the network encodes an entanglement structure in the state.  Taking inspiration from the application of tensor networks in quantum physics as low-rank approximations of many-body ground states, tensor network factorizations of the model weights will be used to constrain the correlation structure between the tensor elements $\{\alpha_k n_k l_k\}$ of the input descriptor.  We will find that this acts as an implicit method to regularize the model fitting \cite{Novikov:arxiv.2015}.

Since there is no inherent geometric relationship between the positions of the descriptor indices, we consider factorizations which preserve their order.  To maintain generality, we will primarily examine factorizations built from matrix product states (MPS),

	\be
		| \psi^{(N)} \rangle = \sum_{s_1, \dotsc, s_N} a_1^{s_1} \dotsm a_N^{s_N} | s_1 \dotsm s_N \rangle \ ,
	\label{mps}
	\ee

\begin{figure}[!h]
	\centering
		\includegraphics[scale=0.25]{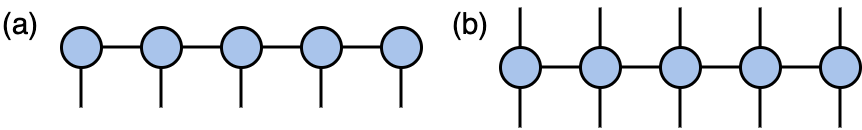}
		\caption{Tensor network representations of (a) a matrix product state (MPS) and (b) a matrix product operator (MPO).}
	\label{mps-mpo}
	\end{figure}

\nn
and matrix product operators (MPO),

	\be
		\hat{T}^{(N)} = \sum_{\substack{s_1,\dotsc, s_{N} \\ s'_1, \dotsc, s'_N}} M_1^{s'_1 s_1} \dotsm M_N^{s'_N s_N} | s'_1 \dotsm s'_N \rangle \langle s_1 \dotsm s_N |
	\label{mpo}
	\ee
\nn
where $a_k^{s_k}$ and $M_k^{s'_k s_k}$ are $\chi_k \times \chi_{k+1}$ matrices, which constitute the learnable model parameters.  Note that because the input descriptors and target properties are real-valued, the tensors $a_k$ and $M_k$ will also be real-valued.  The internal bond dimensions $\chi_k$ are often referred to as virtual dimensions, and the physical indices $s_k$ denote either a descriptor index $\{\alpha_k, n_k, l_k\}$ or a generic, internal vertical bond.  MPS and MPO tensor networks are shown in Figure \ref{mps-mpo} for the case of finite virtual boundary conditions, $\chi_1 = \chi_N = 1$.  For simplicity, we will set all virtual dimensions to be equal, $\chi_k = \chi$, except for the boundary edges.  As discussed further in Section \ref{tn-benchmark}, the virtual dimensions impose upper bounds on the bipartite entanglement entropies of the state $|\psi^{(\nu)} \rangle$.  

\begin{figure}[!h]
	\centering
		\includegraphics[scale=0.25]{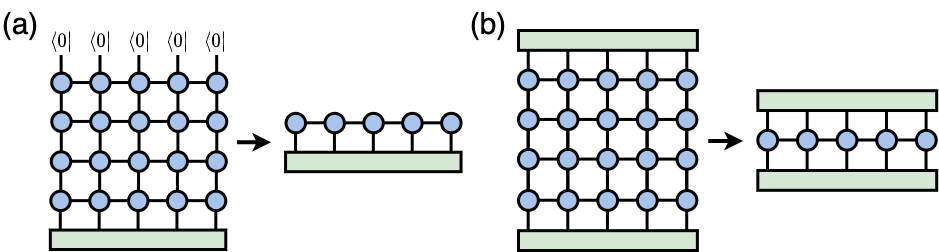}
		\caption{The (a) MPS and (b) MPO models used in this work viewed as the contraction of several MPOs.  In particular, the upper layer on the LHS of the MPS model (a) defines an embedding of an MPS as an MPO.  Learnable model weights are shown in blue, while input descriptors are shown in green.}
	\label{mps-mpo_models}
	\end{figure}

It is useful to note that the sequential application of several MPOs to an arbitrary product state $|0\rangle^{\otimes N}$ is equivalent to evolving the state by a finite-depth, variational quantum circuit.  If each MPO possesses bond dimension $\chi$, then contraction of $n_d$ MPOs with an arbitrary product state yields an MPS with bond dimension $\chi^{n_d}$ (Figure \ref{mps-mpo_models}a).  This can be an efficient way to achieve an MPS with large effective bond dimensions while mitigating the growth in the number of parameters.  For the cases studied in this work, the bond dimensions required to achieve accurate models remain relatively small, so we take as our starting ansatz a single MPS whose bond dimensions are treated as hyperparameters.  Furthermore, since the MPS tensor elements are real-valued, we optimize them directly, rather than relying on unitary embeddings, using classical (e.g., stochastic gradient descent (SGD)) rather than quantum algorithms.  We will return to this tensor network / quantum circuit correspondence in our discussion in Section \ref{tn-benchmark}.  To make use of high-performance automatic differentiation and back propagation algorithms commonly employed in deep learning, the bond dimensions remain fixed during training.  However, performing local tensor updates based on the density matrix renormalization group (DMRG) \cite{Stoudenmire:arxiv.2016, Schollwock:AnnPhys.2011}, which adaptively evolve the bond dimensions, may provide an interesting alternative training method.

While we have focused so far on the construction of learnable states $|\psi^{(\nu)}\rangle$, we can alternatively formulate the task of learning scalar contributions to the target property $V_i$ in terms of an expectation value of an MPO, $\langle B_i^{(\nu)} | \hat{T}^{(3\nu)}  | B_i^{(\nu)} \rangle$, with respect to the invariant basis states $| B_i^{(\nu)} \rangle$.  This can be thought of as a relaxed version of the Born rule, $p(B) = |\langle B | \psi \rangle|^2$, where the pure state density matrix, $|\psi \rangle \langle \psi|$, is replaced by a general mixed state, described by a matrix product density operator (MPDO), $\hat{\rho}$.  Indeed, it was shown in \cite{Glasser:arxiv.2019} within the context of probabilistic graphical models that locally-purified approximations of MPDOs can represent a larger class of non-negative tensors than an MPS model of equivalent rank.  Because we are concerned with the prediction of some target scalar which is not necessarily a probability $p(B)$, we do not enforce non-negativity on the elements of the MPO, $\hat{T}$.  Nonetheless, we will provide empirical evidence in Section \ref{tn-benchmark} that a similar rank-efficiency relationship may hold for regression tasks when measured against their performance on unseen data.  Again, while we could in general train a sequence of MPOs, akin to mapping an invariant basis state back to itself \footnote{This is similar to the treatment of thermal states as path integrals supported on a compact manifold \cite{Evenbly:prl.2015}.}, we will instead consider a single MPO with adjustable bond dimensions (Figure \ref{mps-mpo_models}b). 

	\begin{figure}[!h]
	\centering
		\includegraphics[scale=0.25]{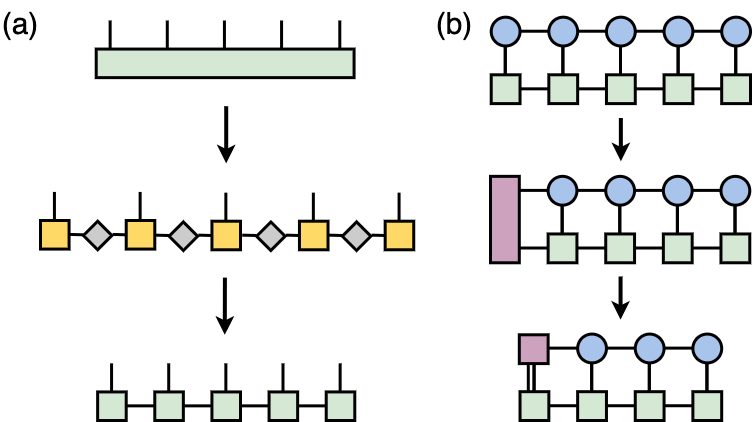}
		\caption{(a) A MPS factorization of a dense tensor can be constructed by the sequential application of singular value decompositions between groups of edges.  The grey diamonds are matrices of singular values which are subsequently contracted with their neighboring tensors.  (b) An efficient contraction order for an MPS model with factorized inputs, where the current contracted tensor is shown in purple.}
	\label{mps-decomp}
	\end{figure}

As the order $\nu$ and the dimensions of the individual indices of the dense descriptor tensor increase, its contraction with the variational states $| \psi^{(\nu)} \rangle$ and operators $\hat{T}^{(3\nu)}$ can become computationally demanding.  However, this issue can be alleviated by analogously constructing low-rank, MPS approximations of the states $| B_i^{(\nu)} \rangle$.  A standard procedure \cite{Schollwock:AnnPhys.2011, Orus:AnnPhys.2014, Bridgeman.JPA_MT.2017, Orus:nrp.2019, Cirac:RMP.2021}, shown in Figure \ref{mps-decomp}a, is to recursively take singular value decompositions (SVDs) between neighboring pairs of edges, discard a subset of the lowest singular values (SV), and contract the corresponding SV matrices with their neighboring tensors.  The virtual bond dimensions of the subsequent MPS is equal to the number of singular values retained at each edge (i.e., the Schmidt rank), which controls the accuracy of the approximation.  For small enough bond dimensions, this constitutes a kind of sparsification procedure on the inputs, and we find in practice (cf., Section \ref{tn:compression}) that model accuracy can be maintained with surprisingly small bonds.  With this MPS decomposition, a more efficient contraction scheme (e.g., Figure \ref{mps-decomp}b), can be implemented.

%----------------------------------------------------------------------------------------
\subsection{Relationship to other methods} \label{tn:other_methods}

In practice, a balance must be sought between the computational efficiency of the model and the number of input features required to accurately represent the atomic environment.  A common simplification is to restrict the expansion (\ref{ACE}) to many-body descriptors of relatively low order.  This is, for instance, the strategy employed in Behler-Parrinello (BP) neural network potentials \cite{Behler:prl.2007}, which use 1-, 2- and 3-body symmetry functions.  It is often the case that features derived from order $\nu \le 3$ expansions are able to differentiate the relevant structural characteristics in a given sample, although certain counterexamples exist \cite{Pozdnyakov:prl.2020}.  Furthermore, as noted in Section \ref{so3-reps}, higher-order descriptors contain products of lower order, which enables accurate models to be built on a fixed-order descriptor \cite{Bachmayr:arxiv:2019, Nigam:jcp.2020}.  This is the case for kernel-based models utilizing either the SOAP power spectrum ($\nu=2$) or bispectrum ($\nu=3$). 

While the expansion (\ref{ACE}) motivates the systematic introduction of SO(3)-invariant descriptors, one is not limited to a linear model for the prediction of some atomic property $V_i$.  Indeed, kernel methods based on SOAP features and neural networks using ACSFs incorporate general forms of nonlinearity, whether through the choice of covariance kernel $K(\mbf{x}, \mbf{x}') $ in the former, or the structure of the learning architecture in the later.  It is worth dwelling on some essential aspects of kernel-based and neural network-based methods, as they will serve as further motivation for the tensor network methods introduced above.

In supervised learning, where the goal is to find a function $f(\mbf{x})$ which approximates a target quantity $y \approx f(\mbf{x})$, a nonlinear function on the inputs $x_i$ can be constructed via a mapping $\varphi$ into a higher-dimensional \textit{feature space}.  Such a mapping allows the function $f$ to be formulated as a linear model on the feature space, but the computational complexity of working directly with very large feature vectors $\varphi(\mbf{x})$ often prohibits their explicit implementation in machine learning tasks.  However, since linear models constructed in a feature space can be rewritten in terms of an inner product $\langle \varphi(\mbf{x}) | \varphi(\mbf{x}') \rangle$ on that space, Mercer's theorem allows one to replace the explicit feature map with a positive, semidefinite kernel function $K(\mbf{x}, \mbf{x}') = \langle \varphi(\mbf{x}) | \varphi(\mbf{x}') \rangle$.  This so-called \textit{kernel trick} and the representer theorem leads to a model of the form $f(\mbf{x}) = \sum_a w_a K(\mbf{x}, \mbf{x}_a)$, where the sum runs over a reference set of training instances.  Note that the cost to evaluate the model scales linearly with the size of the training set, and thus smaller training sets may generalize poorly to out-of-sample data.  Furthermore, the cost for training the model scales like $\mcl{O}(N^3)$ for $N$ training instances.

\begin{figure}[!h]
	\centering
		\includegraphics[scale=0.25]{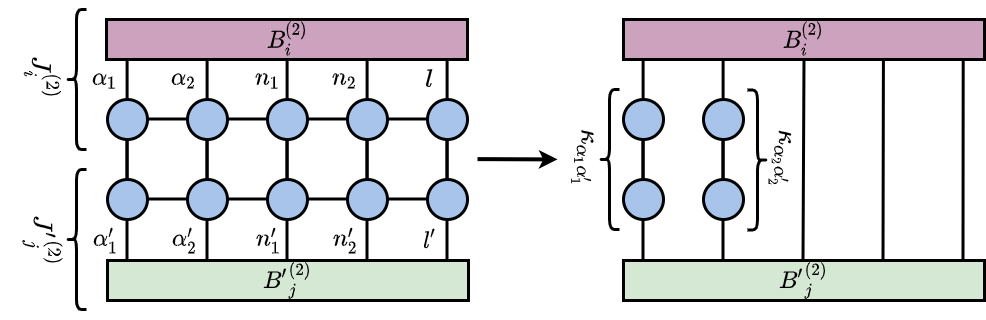}
		\caption{A low-rank approximation of a generalized structural kernel can be represented in terms of MPOs.  For the appropriate arrangement of physical bonds, this contains the class of alchemical kernels shown at right.}
	\label{soap_ker}
	\end{figure}

An advantage of the expansion (\ref{irrep_decomp}) of the local density in the basis of SO(3) irreps is that it comes equipped with a natural inner product, which induces an inner product on the SO(3)-invariant tensor product states $| \sigma_i^{\otimes \nu} \rangle_{g}$.  Under this induced inner product, the feature space can be identified with the descriptor space.  As a tensor network, the associated kernel function is given by contracting the open edges of the SO(3)-invariant descriptors $B_i^{(\nu)}$ and  ${B'}_j^{(\nu)}$ (e.g., Figure \ref{soap_ker}).  Note that in kernel-based methods, pairs of local environments in the kernel function come from different structures, which we distinguish using a prime on the descriptors.  The (smoothed) overlap $k\left(B_i^{(\nu)}, {B'}_j^{(\nu)}\right) \coloneqq \langle B_i^{(\nu)} | {B'}_j^{(\nu)} \rangle$ quantifies the similarity between atomic environments, and raising this kernel to an integer power $\zeta$ enhances its sensitivity to the differences between those environments.  Up to a normalization of the descriptors, these overlaps yield the class of SOAP kernels upon summing over all sites.  Introducing a power $\zeta$ to the overlaps is equivalent to taking a $\zeta$-order tensor product of the invariant basis state, $| B_i^{(\nu)} \rangle^{\otimes \zeta}$, which increases the effective many-body character of the state \cite{Willatt:jcp.2019}.  We will return to this idea when discussing generalizations of our method below.  Note that these $(\nu \cdot \zeta)$-order states do not span the entirety of their SO(3)-invariant tensor product space but instead correspond to states with trivial intermediate irreps (cf. Figure \ref{ladder_of_descriptors}).  

\begin{figure}[!h]
	\centering
		\includegraphics[scale=0.25]{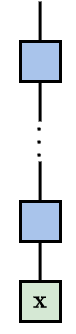}
		\caption{A dense neural network as a simple tensor network, where the internal bond dimensions are determined by the number of nodes in each hidden layer, and the layer-wise activation functions and biases are implicit.}
	\label{ann_tn}
	\end{figure}

A dense (fully-connected, feed-forward) neural network (NN) model approximates the predictor function $f(\mbf{x}) = \mcl{F}\mcl{L} \dotsm \mcl{F}\mcl{L}(\mbf{x})$ by an alternating composition of affine maps $\mcl{L}(\mbf{x}) = w \mbf{x} + \mbf{b}$ and nonlinear activation functions $\mcl{F}(\cdot)$ applied element-wise to their input vectors.  A dense neural network possesses a relatively simple tensor network representation (Figure \ref{ann_tn}), in which the linear weight matrices $w$ are shown as 2-index tensors while the biases $\mbf{b}$ and activation functions remain implicit.  In the case where the SO(3)-invariant descriptor space is taken to be the input feature space and the activation functions and biases are chosen to be trivial, this neural network architecture provides a tensor network factorization of the model weights $w_i^{(\nu)}$ in the ACE expansion (\ref{ACE}), where the internal bond dimensions in the tensor network correspond to the number of nodes in the hidden layers of the neural network, and the output (open) bond has dimension 1 when the target property is a scalar.  Restoring the biases and nonlinear activation functions thus generalizes the linear maps $w_i^{(\nu)}$.  Since the descriptors are treated as input feature vectors with a multi-index $(\alpha_1n_1l_1\dotsm)$, the input bond dimension suffers from the same exponential scaling with $\nu$ as the original ACE expansion.  For deeper neural networks with many hidden nodes, this can lead to an exceptionally large number of model weights and a potential risk of overfitting when the number of training samples is small.

At this point we can draw some connections between the MPO/MPS model in Figure \ref{mps-mpo_models}a and the neural network- and kernel-based methods discussed above.  A sequence of $n_d$ MPOs applied to the input descriptors can be viewed as a particular factorization of the weight matrices $w$ in the $n_d$ layers of a dense neural network.  For sufficiently small virtual bond dimensions, this can lead to a substantial reduction in the number of model parameters \cite{Novikov:arxiv.2015, Gao:prr.2020}.  Hence, the tensor network structure can be understood as a form of regularization on the otherwise dense model weights, where the chosen factorization scheme can enforce a degree of sparsity in the parameters.  Moreover, a single MPO applied to an input state $| B_i^{(\nu)} \rangle$ constitutes a mapping from the symmetry-adapted, atom-centered Hilbert space to a potentially lower dimensional space, $| B_i^{(\nu)} \rangle \ra | J_i^{(\nu)} \rangle = \hat{T}^{(3\nu)} | B_i^{(\nu)} \rangle$.  Note that since this operator acts on the reduced tensor elements $B^{(\nu)}$ of the SO(3)-invariant subspace, it commutes with the action of the SO(3) rotation group.

The overlap $\langle J_i^{(\nu)} | {J'}_j^{(\nu)} \rangle$ is equivalent to a generalized kernel function which couples the structure of the atomic environments to the chemical species within them.  This constitutes a very general form of low-rank approximation for the ``nonfactorizable operators" discussed in \cite{Willatt:jcp.2019}.  Indeed, this overlap contains the class of so-called ``alchemical" kernels built from the SOAP power spectrum (Figure \ref{soap_ker}), which have been shown to improve the accuracy of structural kernel-based methods \cite{De:pccp.2016, Willatt:pccp.2018}.  Adopting the notation in \cite{Willatt:jcp.2019, Willatt:pccp.2018, De:pccp.2016}, we see that the alchemical couplings, $\kappa_{\alpha \alpha'}$, can be decomposed in a lower-dimensional elemental basis $| s \rangle$, $\kappa_{\alpha \alpha'} = \sum_{s} u_{\alpha s} u_{s \alpha'}$.  In previous applications of these generalized kernel methods, the values of these couplings were either chosen explicitly, for instance by incorporating physical intuition \cite{De:pccp.2016, Artrith:prb.2017}, or learned via additional feature selection steps prior to training the model \cite{Willatt:pccp.2018, Imbalzano:jcp.2018}.  Alternatively, by considering general tensor network factorizations of the the weights $w_i^{(\nu)}$, one can work directly with the atom-centered Hilbert space rather than with distinct pairs of local environments.  The above discussion highlights the fact that the choice of network topology imposes a kind of inductive bias on the model which may be used as a way to prioritize a certain kind of solution (e.g., an alchemical one) for the predictor $f(\mbf{x})$.

So far, our construction has focused on tensor decompositions of the weights $w_i^{(\nu)}$ as multilinear maps.  However, as mentioned above, there may be some advantage to introducing nonlinearity into the model architecture.  This could be accomplished analogously to neural networks by introducing element-wise, nonlinear activation functions at every level.  Alternatively, following the work of Stoudenmire and Schwab \cite{Stoudenmire:arxiv.2016}, we could construct an explicit higher-dimensional embedding via a feature map $\varphi$ and work directly with tensor network factorizations of linear weights $\omega$ on the embedding space, $f(x) = \omega \cdot \varphi(\mbf{x})$, without, as with kernel methods, passing to the dual vector space.  A convenient embedding is formed by a (unentangled) product state,

	\be
		\varphi(x) = \varphi^{s_1}(x_1) \otimes \varphi^{s_2}(x_2) \otimes \dotsm \otimes \varphi^{s_N}(x_N) \ ,
	\ee

\begin{figure}[!h]
	\centering
		\includegraphics[scale=0.25]{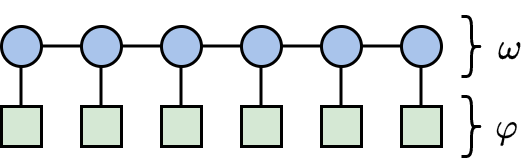}
		\caption{Stoudenmire and Schwab MPS model \cite{Stoudenmire:arxiv.2016}, where each input component is explicitly mapped to an unentangled product state, $\varphi$.}
	\label{ss_mps_model}
	\end{figure}

\nn
where each element of the input vector $\mbf{x}$ (i.e., the vectorization and standardization of the input descriptor $B^{(\nu)}$) is encoded by a local feature map $\varphi^{s_i}(x_i)$.  We found that in practice a simple linear embedding $\varphi^{s_i}(x_i) = \left[ 1 - x_i, x_i \right]^{\intercal}$ performed well on vectorized atom-centered descriptors, outperforming the family of spin-coherent embeddings proposed in \cite{Stoudenmire:arxiv.2016}.  Because $\varphi$ maps a $d$-dimensional input vector to a feature space of dimension $2^d$, one is limited to tensor network factorizations of the weights that can be efficiently contracted.  As in the original proposal, the simplest choice corresponds to a matrix product state (Figure \ref{ss_mps_model}).  Still, to obtain an accurate model, one is often left with a large number of learnable parameters.  Moreover, we have found that TN models built directly on the descriptor space outperform models with an additional feature map, while requiring fewer parameters.

Another interesting possibility is to mimic the construction of nonlinear structural kernels by forming $\zeta$-order tensor products of the symmetrized descriptors $B^{(\nu)}$:

	\be
		B^{(\nu)} \ra B^{(\nu)} \otimes B^{(\nu)} \ra \cdots \ra \left(B^{(\nu)}\right)^{\otimes \zeta}
	\label{copy_ops}
	\ee
	
\nn
As inputs to a NN or TN model, this construction entails reuse of information in the subsequent network.  Information reuse is a common characteristic of modern NN architectures, such as convolutional (CNN) and recurrent (RNN) neural networks, where several trainable filters act on the same input or subsets of input \cite{Glasser:IEEEa.2020}.  Recent work \cite{Levine:arxiv.2017, Levine:prl.2019} has shown that convolutional (CAC) and recurrent (RAC) arithmetic circuits, which share fundamental architectural components like overlapping filters and pooling with conventional CNNs and RNNs, can be mapped to so-called \textit{generalized} tensor networks \cite{Biamonte:aipa.2011, Glasser:IEEEa.2020} by introducing local copy operations into the tensor algebra \footnote{The general copying of vectors and tensors by a standard tensor network on a finite-sized Hilbert space is forbidden by a no-cloning theorem.}.  It was shown in \cite{Levine:arxiv.2017, Levine:prl.2019} that a deep CAC with a local pooling scheme maps to a hierarchical, tree-structured tensor network and, for maximally overlapping filters, can support volume-law scaling of the entanglement entropy when modeling the amplitudes of many-body quantum states.  Moreover, the entanglement capacity \cite{Cui:jmp.2016} of the CAC tensor network was shown empirically to strongly influence the inductive bias of supervised image classification with a CNN \cite{Levine:arxiv.2017}.

In a similar vein, the MPO model introduced above can be viewed as a specific choice of learnable architecture acting on $\zeta=2$ copies of the input descriptor, $T^{(3 \nu)} \left( | B^{(\nu)} \rangle^{\otimes 2} \right)$.  Since copy operations (\ref{copy_ops}) applied to the input descriptors increase their effective many-body character \cite{Willatt:jcp.2019}, this method provides a potential route toward capturing higher-order correlations with lower order features.

%----------------------------------------------------------------------------------------
%----------------------------------------------------------------------------------------
\section{Numerical Benchmarks} \label{tn-benchmark}
To validate the methods described in Section \ref{tensor-facts}, we will focus on a typical learning task encountered in materials property prediction using a widely available dataset (NMD18 \cite{Sutton:npj-cm.2019}) of 3000 transparent conducting oxide (TCO) alloys (Al$_x$Ga$_y$In$_{z}$)$_2$O$_3$, a class of wide bandgap materials with high technological applicability in optoelectronic devices.  The target property is the DFT mixing energy per cation (referred to as the formation energy in \cite{Sutton:npj-cm.2019}) for a given configuration, referenced to the compositional endpoints of the material,

	\begin{align}
		\Delta H_\mrm{mix} &[(\mrm{Al}_x \mrm{Ga}_y \mrm{In}_{z})_2 \mrm{O}_3] = E[(\mrm{Al}_x \mrm{Ga}_y \mrm{In}_{z})_2 \mrm{O}_3] \nonumber \\
		&- x E[\mrm{Al}_2 \mrm{O}_3] - y E[\mrm{Ga}_2 \mrm{O}_3] - z E[\mrm{In}_2 \mrm{O}_3] \ .
	\label{form_enthalpy}
	\end{align}

\nn
Here $x + y + z = 1$, and $E[c]$ is the ground state DFT energy per cation for the given compound $c$.  The mixing energy characterizes the (zero-temperature) stability of the alloy configuration relative to the single cation phases.  The underlying crystalline lattices of the alloy configurations span six distinct space groups (C/2\textit{m}, P\textit{na}$2_1$, \textit{R}$\overline{3}$\textit{c}, P$6_3$/\textit{mmc}, I\textit{a}$\overline{3}$, and \textit{Fd}$\overline{3}$\textit{m}) and the number of sites in each structure can vary in integer multiples of the number of primitive cell sites.  The reference compositional endpoints, however, are fixed to their ground-state lattices, \textit{R}$\overline{3}$\textit{c} for Al$_2$O$_3$, C/2\textit{m} for Ga$_2$O$_3$, and I\textit{a}$\overline{3}$ for In$_2$O$_3$.  The inclusion of multiple lattice symmetries is difficult to handle with more conventional atomistic modeling methods, such as the standard cluster expansion (CE) \cite{Sutton:npj-cm.2019, Nyshadham:npj-cm.2019}.  Moreover, allowing for chemically ordered/disordered sublattices presents challenges to deep end-to-end learning methods, where very large datasets are often required to obtain sufficient accuracy \cite{Dunn:npj_cm.2020, Zuo:jpca.2020, Chen:ncs.2021}.  Depending on the application and the computational cost of the first-principles calculations, generating a large dataset can be prohibitively expensive.  It is therefore useful to understand how the accuracy of these methods scales with the size of the training dataset.

To further emulate working in a data-constrained environment, we do not assume prior knowledge of the fully relaxed atomic positions and lattice constants.  Instead, the input structures retain their ideal lattice positions, while the lattice vectors are simply scaled according to Vegard's law.  The target properties, however, are calculated from the relaxed geometries.  This constitutes a particularly challenging scenario.  Indeed, it was previously found that learning a mapping from unrelaxed structures to ground-state energies is generally more difficult than using relaxed input structures, at least at the level of kernel-based learning \cite{Langer:arxiv.2020}.  Moreover, the atomic descriptor was not found to be the limiting factor, but rather prediction error was dominated by implicit noise in the underlying set of atomic structures \cite{Langer:arxiv.2020}.

To construct a global descriptor suitable for the prediction of the global property $\Delta H_\mrm{mix}$, we take an average over the local, atom-centered descriptors of the structure,

	\be
		\overline{B}^{(\nu)} = \frac{1}{N} \sum_i^N B_i^{(\nu)} \ .
	\label{avg_descriptor}
	\ee

\nn
This is a special case of the general prescription of calculating $\Delta H_\mrm{mix}$ as a sum over (possibly distinct) contributions $H^{(\nu)}_i$ from $N$ local environments, where for (\ref{avg_descriptor}) all local $\nu$-order contributions are contracted with the same weight tensor $w^{(\nu)}$.  To simplify the comparison between different machine learning methods and architectures, we choose the SOAP power spectrum $B^{(2)}$ with fixed radial and angular momentum cutoffs ($R_c = 6$ $\mrm{\AA}$, $n_\mrm{max} = 4$, $l_\mrm{max} = 3$) to be the input descriptor.  This also allows us to compare our results to those reported in \cite{Sutton:npj-cm.2019, Langer:arxiv.2020}, where the former tested the performance of both a deep NN with SOAP feature vectors and Gaussian process regression (GPR) with the SOAP kernel ($\zeta=2$), and the latter employed kernel ridge regression (KRR) using a Gaussian / radial basis function (RBF) kernel.  Thus, in addition to systematically evaluating various tensor network architectures, we provide consistent benchmarks with respect to the performance of GPR with SOAP and RBF kernels, as well as the deep NN architecture described in \cite{Sutton:npj-cm.2019}.

%----------------------------------------------------------------------------------------
\subsection{Learning curves}
We quantified the predictive performance of the MPS and MPO models, as well as GPR and NN models, using stratified k-fold cross-validation.  In particular, we considered 10 splits of the NMD18 dataset into a testing set of 600 structures and a remaining pool for training and validation.  Of this remaining pool of structures, 10 subsets of a fixed size were chosen with consistent distributions of volume, composition and energies, following \cite{Langer:arxiv.2020}, from which we constructed 80-20 splits into training and validation sets.  Prediction errors were measured on the testing sets, which remain untouched during model training, while validation sets were used for hyperparameter tuning.

	\begin{figure}[!t]
	\centering
		\includegraphics[scale=0.35]{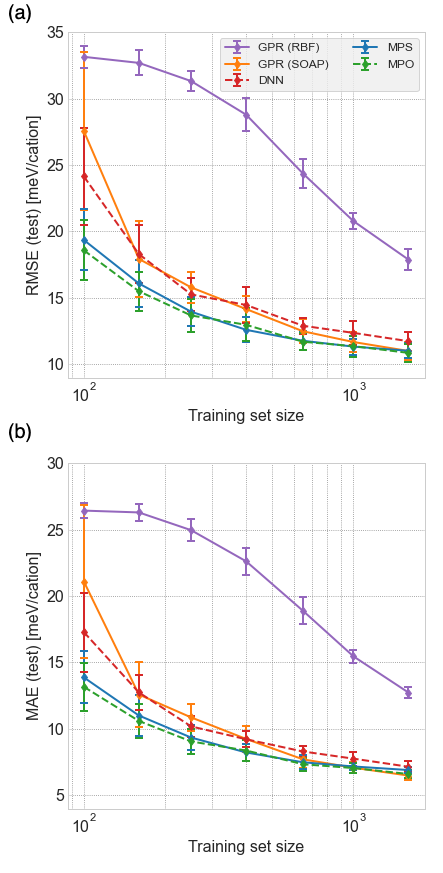}
		\caption{Test RMSE (a) and MAE (b) learning curves on the NMD18 dataset.  MPS and MPO models are compared to a fully-connected, deep neural network (DNN) and Gaussian process regression (GPR) with RBF and SOAP kernels.}
	\label{learning_curves}
	\end{figure}

Figure \ref{learning_curves} compares the average and standard deviations of the root mean square error (RMSE) and the mean absolute error (MAE) of the model predictions over the testing splits, as a function of the training set size.  First, while it shows consistently small variance, GPR with an RBF kernel performs the worst among the models considered, likely due to over-localization of the atomic representations in the embedding feature space.  By contrast, GPR with the SOAP kernel exhibits higher sensitivity to the differences between training structures, improving substantially as the number of reference structures increases.  As observed in \cite{Sutton:npj-cm.2019}, we find that the DNN model and GPR with SOAP kernel perform similarly across training sets.  However, the number of model weights in the DNN architecture remains fixed, while the number of weights increases with the training set size in the kernel methods.  Importantly, the DNN is incrementally improvable with the acquisition of new data, while the kernel-based models must be retrained on the whole dataset.  Moreover, for much larger datasets, the computational cost of kernel-based methods becomes prohibitively expensive.

The MPS and MPO models maintain strong generalizability even for very small ($\mcl{O}(10^2)$ structures) training sets.  In this data-constrained regime, they outperform the other models considered, while converging with the DNN and GPR (SOAP) models at the largest training set size.  The advantage of the TN models is particularly notable in the RMSE, which is more sensitive to the presence of outliers than the MAE.  As with the DNN model, the TN models are systematically improvable with new data without requiring retraining on the full dataset.  Unlike the DNN model, they require substantially fewer parameters ($\mcl{O}(10^3)$ compared to $\mcl{O}(10^6)$) to achieve similar accuracy.  This parametric efficiency may, however, be related to the observed onset of saturation in the model accuracy, particularly for the MAE of the MPS model, as the training set size increases.  Yet, similar saturation is also apparent in the DNN model.  Moreover, the MPO model marginally outperforms the MPS model, despite possessing a lower virtual bond dimension, and appears less prone to plateaued accuracies for increasing training set sizes.  As we show below, the accuracies of the TN models do not suffer from increasing the bond dimensions for fixed datasets, and thus we expect that this saturation can be overcome by simply increasing the number of parameters.  It is worth noting that we have additionally tested deeper TN factorizations with stronger entanglement scaling, specifically tree tensor networks (TTN) \cite{Shi:prl.2006} and the Multi-Scale Entanglement Renormalization Ansatz (MERA) \cite{Vidal:prl.2007} (Figure \ref{various_tns}), and found that they produce accuracies comparable to the MPS model.  They are, however, more computationally expensive to contract.  Finally, the computational cost of the TN models scales linearly with the number of samples, as opposed to the quadratic scaling of the kernel-based methods.  Thus, we expect the application of these models to be practical across a broad range of dataset sizes.

%----------------------------------------------------------------------------------------
\subsection{Entanglement measures}
We have seen above that the MPS and MPO models can yield expressive supervised learning models even with a small number of training samples.  To what extent does the choice of TN architecture influence this apparent inductive bias?  Since the topology and bond dimensions of the TN constrain the local entanglement structure of the state $| \psi^{(\nu)} \rangle$, it is reasonable to expect that the entanglement in the learned TN parameters captures underlying correlations in the input data.  Indeed, similar reasoning has been used to justify a priori the choice of a given TN machine learning architecture by characterizing the spatial scaling of the entanglement entropy \cite{Levine:arxiv.2017, Martyn:arxiv.2020} or the mutual information \cite{Convy:arxiv.2021, Lu.arxiv.2021} for bipartitions of the input features.  These studies were based primarily on the analysis of image data, which possess a precise notion of spatial arrangement.  Similar analyses could be performed by measuring correlations between individual features $x_i$ or their product state embedding $\varphi(x)$, in the original spirit of Stoudenmire and Schwab \cite{Stoudenmire:arxiv.2016}.  However, since our TN models operate directly on the tensor structure of the input descriptors, the number of available partitions of the tensor indices is small.  Hence, an analysis and interpretation of their scaling with subsystem size is limited, except perhaps at large order $\nu$.  Instead, in this section, we scrutinize the entanglement learned by the models themselves, conditioned on the target property.

Let us recall that the entanglement entropy is a zero-temperature quantum analogue of the classical Shannon entropy.  We will adhere to common practice by taking the entanglement entropy to mean specifically the bipartite, von Neumann entropy,

	\be
		S_\mrm{vN}(\rho_\mcl{A}) \coloneqq -\Tr \left( \rho_\mcl{A} \log \rho_\mcl{A} \right) \ ,
	\label{vn_ee}
	\ee
	
\nn
where $\rho_\mcl{A}(| \psi \rangle)$ is the reduced density matrix of a subsystem $\mcl{A}$ of dimension $d_\mcl{A}$, defined by tracing out the degrees of freedom of a complementary subsystem $\mcl{B}$ in a pure state $| \psi \rangle$,

	\be
		\rho_\mcl{A}(| \psi \rangle) = \Tr_\mcl{B}\left( | \psi \rangle \langle \psi | \right) \ .
	\label{schmidt_decomp}
	\ee

\nn
The entanglement entropy can be computed from the coefficients of a Schmidt decomposition of the pure state $| \psi \rangle$,

	\be
		| \psi \rangle = \sum_{k=0}^{\min(d_\mcl{A},d_\mcl{B})} \lambda_k |\psi^{k}_\mcl{A} \rangle | \psi^{k}_\mcl{B} \rangle \ .
	\ee

\nn
The Schmidt spectra $\lambda_k$ are the singular values of the of matrix $C_{\mcl{A}\mcl{B}}$, where $| \psi \rangle = \sum_{\psi_\mcl{A}, \psi_\mcl{B}} C_{\mcl{A}\mcl{B}} |\psi_\mcl{A} \rangle | \psi_\mcl{B} \rangle$.  Since the eigenvalues of the reduced density matrix $\rho_\mcl{A}$ are $\lambda_k^2$, the entanglement entropy reduces to $S_\mrm{vN}(\rho_\mcl{A}) = \sum_k - \lambda_k^2 \log(\lambda_k^2)$.

\begin{figure}[!h]
	\centering
		\includegraphics[scale=0.25]{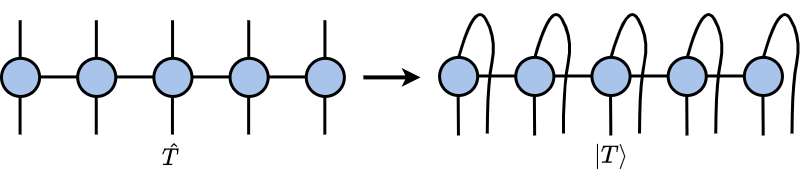}
		\caption{The Choi isomorphism constitutes a vectorization $| T \rangle$ for MPO $\hat{T}$.}
	\label{choi_isom}
	\end{figure}

We can formulate an analogous entanglement entropy for the MPO models (\ref{mpo}) by constructing a canonical purification \cite{vNieuwenburg:pra.2018, Weimer:rmp.2021}.  A vectorization $| T \rangle $ of the MPO $\hat{T}$ follows from the Choi isomorphism \cite{Choi:LAA.1975, Jamiolkowski:RMP.1972, Weimer:rmp.2021} (Figure \ref{choi_isom}), from which an effective pure state density matrix can be defined, $\mcl{Q} = | T \rangle \langle T |$.  We note that under this isomorphism, an MPO with physical dimensions $d$ is mapped to an MPS with physical dimension $d^2$.  Equivalent definitions of the entanglement entropy and Schmidt spectra thus follow from (\ref{vn_ee}) and (\ref{schmidt_decomp}), replacing $\rho$ by $\mcl{Q}$ and $| \psi \rangle$ by $| T \rangle$.

\begin{figure}[!h]
	\centering
		\includegraphics[scale=0.275]{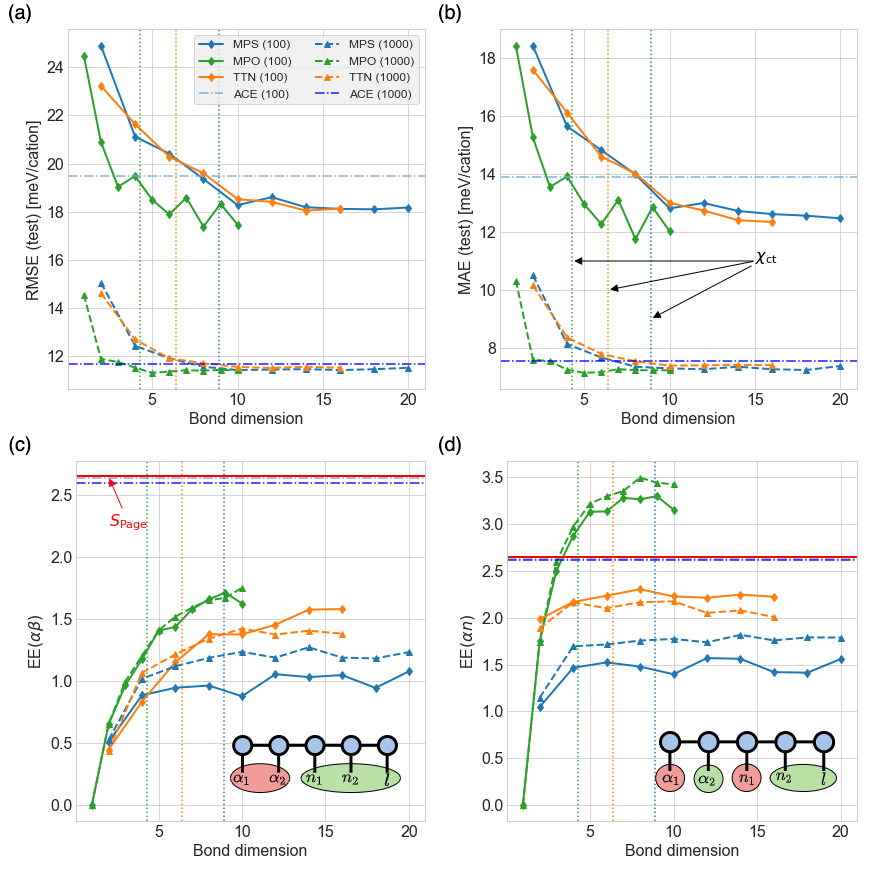}
		\caption{Test RMSE (a) and MAE (b) as a function of the virtual bond dimension used in MPS, MPO and TTN models, for training sets containing either 100 or 1000 structures.  This is compared to the performance of a model using a dense ACE tensor (dashed, horizontal lines), and the corresponding compression thresholds, $\chi_\mrm{ct}$, for each TN model are marked with vertical, dotted lines.  The associated entanglement entropies (c,d) are plotted for different bipartitions (insets; partition $\mcl{A}$ in red, $\mcl{B}$ in green), and the corresponding Page entropy, $S_\mrm{Page}$, is indicated by a solid, red line.  Note that because the MPO model acts on two copies of the input tensor, the effective physical dimensions in each partition (as an MPS under the Choi isomorphism) are larger.  The Page entropy associated with this larger Hilbert space is not shown.}
	\label{ee_error}
	\end{figure}

Figure \ref{ee_error} compares average errors and entanglement entropies of the MPS and MPO models, as well as a TTN model (Figure \ref{various_tns}), as a function of their virtual bond dimensions.  The average is taken over 10 training-validation splits with a fixed testing set, and we test two sizes (100 and 1000 structures) of the training set.  For the entanglement entropies, we consider two bipartitions of the descriptor indices: (1) a contiguous subsystem $\mcl{A}$ consisting of chemical degrees of freedom $\{\alpha_1 \alpha_2\}$ and $\mcl{B}$ containing the structural components $\{n_1 n_2 l \}$, and (2) a noncontiguous subsystem $\mcl{A}$ consisting of a chemical and radial component $\{\alpha_1 n_1\}$ with $\mcl{B}$ containing the remaining degrees of freedom.  These bipartitions are shown as insets in Figures \ref{ee_error}c,d.  For comparison, we also plot the errors and entanglement entropies for the corresponding models using a dense ACE tensor (Figure \ref{ace_tensor}), and we indicate the bond dimension $\chi \le \chi_\mrm{ct}$ for which the number of TN model parameters is less than the dense ACE tensor.  We will refer to this bond dimension $\chi_\mrm{ct}$ as the ``compression threshold."

For the three TN architectures, we generally find that improved test errors are strongly correlated with larger entanglement enabled by higher bond dimensions.  This is particularly true for the smaller training set, which requires larger bond dimensions before reaching minimal errors.  The TN models outperform the ACE models once the bond dimension reaches the compression threshold, that is, once the number of parameters exceeds that of the ACE tensor.  Again, this improvement is more significant in the small training set than in the larger one.  In the larger training set, the test errors of the TN models become competitive with the ACE model well before reaching the compression threshold.  This behavior is somewhat counterintuitive in the context of classical statistical learning, where one expects the generalizability of an overparameterized model to degrade by overfitting on small amounts of training data, a consequence of the so-called \textit{bias-variance tradeoff} \cite{Hastie:esl}.  However, this phenomena is not uncommon in deep neural networks, where models with trainable parameters greatly exceeding the training sample size nonetheless perform well on unseen data \cite{Belkin:arxiv.2018, Nakkiran:arxiv.2019, Nakkiran:arxiv.2020}.

We observe that the average entanglement entropies converge to values well below the theoretical limits set by the min-cuts in the network \cite{Cui:jmp.2016} and that this convergence closely follows the formation of plateaus in test errors.  Similar sub-maximal convergence in the entanglement entropy was also noted in \cite{Gao:prr.2020} for MPO layers applied to the MNIST dataset, so we expect that this is a general phenomenon when the TN model is complex enough for the learning task.  However, the converged entropies in our work appear to be nonuniversal, exhibiting strong dependence on the structure of the network.  Indeed, we generally find that networks which can host stronger entanglement entropy scaling with subsystem size relative to others (for instance, logarithmic versus area law scaling between TTNs and MPSs \cite{Shi:prl.2006, Orus:nrp.2019}) tend to converge to higher values.  Furthermore, for fixed virtual bond dimension, we find that the MPO model outperforms the MPS model and achieves higher entanglement.  This appears consistent with the findings \cite{Glasser:arxiv.2019} of greater representational efficiency of MPDOs versus MPSs in probabilistic graphical models, although, for this work, it is in the context of regression.  Hence, under the same arrangement of virtual bonds, the entanglement entropy seems to provide a consistent measure of the TN model's generalizability.  

Nonetheless, comparisons of the entanglement between trained models with distinct virtual bond topologies are difficult to make.  For instance, the dense ACE tensors perform somewhat worse than the other TN models, but achieve much higher entanglement entropies.  In this case, the entanglement entropies do approach a universal value, that of the Page entropy \cite{Page:prl.1993}, 

	\be
		S_\mrm{Page} = \frac{1-d_\mcl{A}}{2 d_\mcl{B}} + \sum_{k=d_\mcl{B}+1}^{d_\mcl{A}d_\mcl{B}} \frac{1}{k}
	\ee
	
\nn
defined as the average entanglement entropy of a pure state randomly drawn from the entire Hilbert space.  This would not be entirely surprising for an untrained ACE tensor $w^{(\nu)}$ treated as a length $(|\alpha|^\nu |n|^\nu |l|^\nu)$ vector since the individual elements are initialized according to a standard normal distribution $\mcl{N}(0,\sigma^2)$.  However, it is surprising that the high entanglement of this initial state is essentially preserved under SGD with a mean square error (MSE) loss function, at least up until early stopping.  This may provide some explanation for the improved performance of the models with an explicit TN factorization: the entanglement constraints imposed by the network itself help drive the model toward minima in the loss landscape which are characterized by entanglement more finely tuned to the target property.  

It has long been recognized that the entanglement entropy alone is insufficient to fully capture the entanglement of general quantum states and that much richer structure can be found in the full entanglement spectrum $\{\lambda_k^2\}$ or its logarithms $\{\xi_k \coloneqq - \log(\lambda_k^2)\}$ \cite{Li:prl.2008}.  The level spacing statistics between consecutive pairs of eigenvalues have proven useful in characterizing the irreversibility in quantum circuits \cite{Chamon:prl.2014, Shaffer:iop.2014} and emergent entanglement complexity in many-body quantum dynamics \cite{Yang:prb.2017, Zhang:prb.2020, Wiersema:prxq.2020, Iaconis:prxq.2021}.  For entanglement spectra arranged in descending order $ \lambda_{0}^2 \ge  \lambda_{1}^2 \ge \dots$, let $s_k \coloneqq \lambda_{k}^2 - \lambda_{k+1}^2$ denote the $k$th level spacing.  To avoid unfolding the spectrum \cite{Mehta:rmt}, a procedure sensitive to spurious finite size effects, it is common practice \cite{Oganesyan:prb.2007, Pal:prb.2010, Yang:prb.2017, Atas:prl.2013} to study the ratio of consecutive level spacings $r_k = s_k/s_{k+1}$ or the related quantity $\widetilde{r}_k = \min(r_k, 1/r_k)$, which are independent of the level density of states.  It was shown in \cite{Chamon:prl.2014, Shaffer:iop.2014, Yang:prb.2017} that either the distribution of spacings $P(s)$ or their ratios $P(r)$, collectively referred to as the \textit{entanglement spectrum statistics} (ESS), provides a measure of a quantum state's entanglement complexity, defined by the existence of an efficient algorithm that completely disentangles the state.  Complexly entangled states, for which efficient disentangling algorithms fail, were found to feature ESS with Wigner-Dyson (WD) statistics,
	
	\be
		P_\mrm{WD}(r) = \frac{1}{Z_\beta} \frac{(r + r^2)^\beta}{(1 + r + r^2)^{1 + (3/2)\beta}} \ ,
	\ee

\nn
where $Z_\beta$ is a normalization factor, and the Dyson index, $\beta$, specifies one of three Gaussian random matrix ensembles \cite{Atas:prl.2013}.  On the other hand, states which could be efficiently disentangled possessed Poisson-like ESS, $P_\mrm{Poisson}(r) = 1/(1+r)^2$.  An important feature that distinguishes WD from Poisson level statistics is the presence of level repulsion, $P_\mrm{WD}(r \ra 0) \sim r^\beta \ra 0$, which reflects universal statistical correlations between adjacent levels \cite{Mehta:rmt}.  We also note that there appears to be a deep connection between quantum circuits capable of universal computing and ESS \cite{Chamon:prl.2014, Shaffer:iop.2014, Yang:prb.2017, Zhang:prb.2020}, in that a universal gate set gives rise to WD statistics.  However, a recent study \cite{Iaconis:prxq.2021} showed that the converse is not true in general, finding that a class of classically simulatable circuits can prepare states with WD ESS.

It is reasonable to expect the above picture to hold, to some extent, for a randomly initialized MPS.  Indeed, a random MPS, viewed as a unitary embedding \cite{Garnerone:pra.2010, Haferkamp:PRXQ.2021, Liu:arxiv.2021}, forms an approximate 2-design \cite{Harrow:cmp.2009}: as a random quantum circuit, the first and second moments approximate those of a Haar distribution.  Moreover, Haar-distributed random unitary circuits possess WD ESS corresponding to the Gaussian unitary ensemble (GUE), $\beta=2$ \cite{Zhang:prb.2020}, although a precise relationship between ESS and $k$-designs remains an open question \cite{Iaconis:prxq.2021}.  Nonetheless, we will explore the ESS of our MPO model, viewed as an MPS under the Choi isomorphism.  For simplicity, we consider the entanglement spectrum for the $\{\alpha n\}$ bipartition shown in Figure \ref{ee_error}d and analyze its evolution under SGD for a fixed training set.  To obtain sufficient statistics, we collect data over 100 independent runs of 1000 SGD epochs, and we study models with virtual bond dimensions for which the spectrum is either truncated ($\chi = 12$), marginal ($\chi = 16$), or full-rank ($\chi = 20$).  

\begin{figure*}[!t]
	\centering
		\includegraphics[scale=0.25]{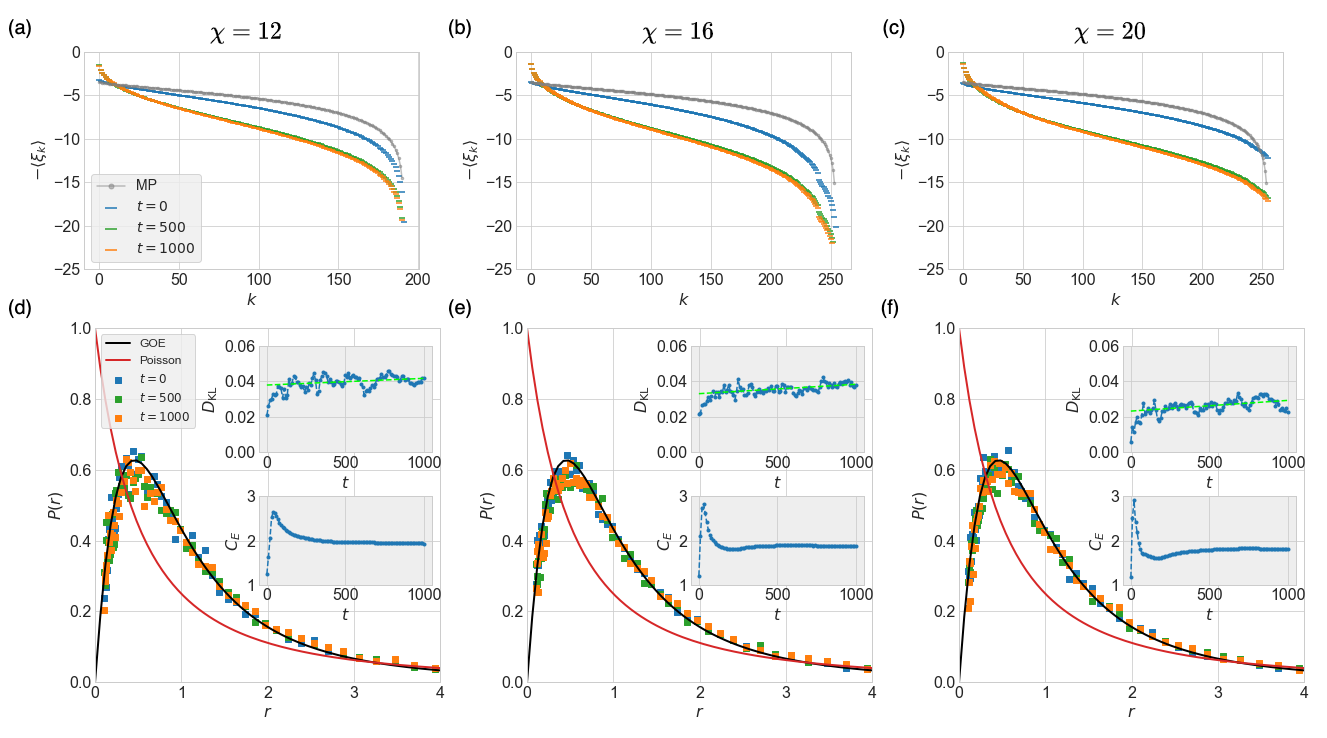}
		\caption{(a-c) The average entanglement spectrum of the MPO model as a function of training step, $t$, compared to the Marchenko-Pastur (MP) law.  (d-f) The distribution $P(r)$ of the ratio $r$ of level spacings in the entanglement spectrum (referred to as the entanglement spectrum statistics (ESS) in the main text) closely follows the Gaussian orthogonal ensemble (GOE).  Under stochastic gradient descent (SGD), the KL divergence (upper insets) exhibits a short, initial regime of fast growth followed by a slow growth regime.  A linear fit to the KL divergence in the slow growth regime (dashed, light blue line) is included to help guide the eye.  The crossover between the fast and slow growth in the observed ESS relative to the GOE, coincides with a divergence in the capacity of entanglement, $C_E$, (lower insets) signaling a phase transition in the entanglement spectrum.}
	\label{es_stats}
	\end{figure*}

Figure \ref{es_stats} shows both the logarithm of the average spectrum $\langle \xi_k \rangle$ and the ESS at different time steps.  First, we find that for all cases the entanglement spectrum deviates from the Marchenko-Pastur (MP) law for Haar-distributed random states sampled from the full Hilbert space \cite{Znidari:jpa_mt.2006, Forrester:rmt}.  Interestingly, close examination of the truncated ($\chi = 12$) model reveals residual MP structure in the tail of the spectrum, similar to the two-component structure discussed in \cite{Yang:prl.2015}, and this MP tail persists under SGD up to a constant shift from normalization.  For all models, upon random initialization of the MPS tensors, the spectrum is relatively flat and thus highly entangled.  The largest changes in the spectrum occur at the earliest stages of training, and the spectrum quickly converges to a configuration in which the local level density has decreased.  We also find that the largest spectra (low $\xi_k$) approximately converge to the same values, regardless of the bond dimension, which is consistent with the prior observation that the model retains high accuracy for smaller bonds.  

At initialization, all three cases display WD ESS corresponding to the Gaussian orthogonal ensemble (GOE), $\beta=1$.  This may not be entirely surprising given the above discussion, although in this case the generating random matrix ensemble is real-valued.  However, as the state evolves under SGD, the ESS retains a substantial degree of its GOE character.  To measure the difference between the GOE ESS, $P_\mrm{GOE}(r)$, and the observed ESS, $P_\mrm{obs}(r)$, we plot the Kullback-Leibler (KL) divergence, $D_\mrm{KL} \left( P_\mrm{GOE} || P_\mrm{obs} \right) = \Tr_r \left[ P_\mrm{GOE} \log\left(P_\mrm{GOE}/P_\mrm{obs} \right) \right]$, in the insets to Figure \ref{ee_error}.  First, we note that the higher starting values of $D_\mrm{KL}$ in the truncated and marginal spectra relative to the full-rank case are likely due to finite-size effects imposed by the bond dimension.  We see that at early times there is rapid growth in the KL divergence followed by a regime in which $D_\mrm{KL}$ increases very slowly, accompanied by oscillations due to the stochasticity of the optimization.  The onset of this slow growth regime prevents the ESS from significantly deviating from the GOE over the course of training.  As we noted above, the regime of rapid growth is accompanied by a decrease in the level density, which essentially stabilizes in the slow growth regime.  

Recalling the level repulsion in the GOE, this expansion of the level density resembles the expansion and re-equilibration of a one-dimensional Coulomb gas following a pertubation of its confining potential.  Specifically, the joint probability distribution of $N$ eigenvalues in the GOE is precisely a Boltzmann-Gibbs factor for a gas of $N$ particles with pairwise, repulsive $-\log(|\lambda_i - \lambda_j|)$ interactions in a quadratic potential \cite{Forrester:rmt}.  This picture suggests that SGD evolves the entanglement spectrum and the ESS by evolving the confining potential of the joint level distribution.  Furthermore, if this potential is driven far enough, the entanglement spectrum should undergo a phase transition.  We can examine whether this is the case by calculating the so-called \textit{capacity of entanglement} \cite{Yao:prl.2010},

	\be
		C_E \coloneqq \langle H_\mrm{ent}^2 \rangle - \langle H_\mrm{ent} \rangle^2 \ ,
	\ee

\nn
where $H_\mrm{ent} \coloneqq - \log \rho_\mcl{A}$ is the \textit{entanglement Hamiltonian} \cite{Li:prl.2008, deBoer:prd.2019}.  The capacity of entanglement measures fluctuations in the entanglement spectrum, analogous to the heat capacity in classical statistical mechanics.  The insets to Figure \ref{ee_error} show the evolution of $C_E$ for the three cases.  We observe that the crossover from the fast to slow growth regimes in the KL divergence coincides with a divergence in the capacity of entanglement, signaling a phase transition in the entanglement spectrum.  Again, this transition is smoothed out in the truncated and marginal cases due to finite-size effects.

The above discussion provides a mechanism through which the model entanglement adapts to the learning task.  Crucially, while the average entanglement spectrum converges to a specific configuration, particularly near its low-$\xi_k$ edge, the ESS maintains universal characteristics of WD statistics, namely level repulsion.  Thus, the learned state retains the entanglement complexity from its random initialization.  This is consistent with its strong generalizability in that the model is expressive enough to capture complex correlations in the high-dimensional feature space conditioned on the target property.  Moreover, this behavior appears to have close connections to overparameterized neural networks \cite{AllenZhu:arxiv.2018, Li:arxiv.2018} and quantum circuits \cite{Wiersema:prxq.2020, Kiani:arxiv.2020}, where SGD applied to sufficiently overparameterized models finds solutions with small generalization error that are nonetheless close to their random initialization.  The transition observed in the entanglement spectrum during training is reminiscent of the dynamical phase transition observed in SGD-based training of deep neural networks \cite{Shwartz:arxiv.2017}.  At early times, model evolution is dominated by the average gradient in the loss function, which rapidly minimizes the model error.  At latter times, model training becomes dominated by stochastic fluctuations in the gradient, introducing diffusive behavior in the model evolution.  This appears to be reflected in the dynamics of the capacity of entanglement, where in the slow growth regime the entanglement spectrum exhibits stronger fluctuations.  In \cite{Shwartz:arxiv.2017}, this diffusion phase was associated with compression of the model's latent representation of the inputs.  We observe similar compressive behavior in the evolution of the entanglement spectrum, where contributions to the entanglement entropy become increasingly dominated by the low-$\xi_k$ edge of the entanglement spectrum.

%----------------------------------------------------------------------------------------
\subsection{Latent space encoding} \label{tn:latent}
As discussed in Section \ref{tn:other_methods}, TN factorizations of the learnable states $|\psi^{(\nu)}\rangle$ can be understood as multilinear maps which internally couple structural and chemical degrees of freedom.  In this section, we explore this idea further by visualizing the latent tensor product spaces learned by a deeper TN architecture (Figure \ref{various_tns}) corresponding to a MERA with fixed virtual bond dimension.  To do so, we use t-distributed stochastic neighbor embedding (t-SNE), a nonlinear dimensional reduction method that preserves the relative locality between points in the dataset \cite{Hinton:nips.2002, vdMaaten:JMLR.2008, Mehta:physrep.2019}.  Again, the MERA network is trained to predict the mixing energy (\ref{form_enthalpy}) on a 1000-structure subset of the full NMD18 dataset, and we employ t-SNE to embed the tensor product spaces mapped by subsequent hidden layers.  

	\begin{figure}[!b]
	\centering
		\includegraphics[scale=0.12]{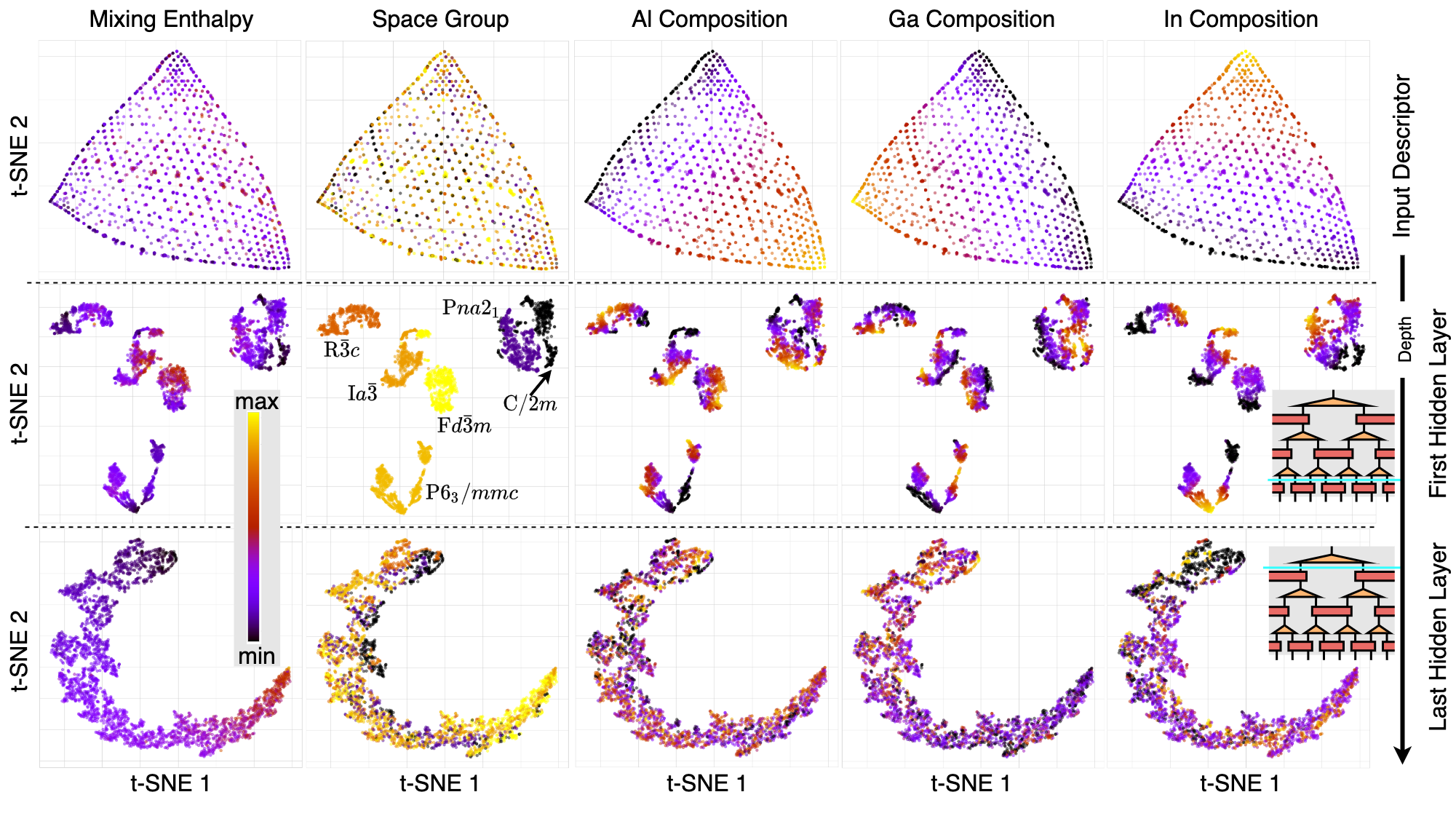}
		\caption{Evolution of t-distributed stochastic neighbor embeddings (t-SNE) of the NMD18 dataset with increasing depth in a MERA tensor network model (insets).  The regularity of the encoding of chemical composition in the input SOAP descriptor is reorganized by the first hidden layer according to the lattice symmetries of the structures.  The features are subsequently coarse-grained by fitting the model to the mixing energy, yielding a quasi-one dimensional encoding ordered by high and low energy.}
	\label{TSNE}
	\end{figure}
	
Figure \ref{TSNE} shows the evolution of the latent space image of each datapoint as function of depth in the MERA network.  Note that we apply t-SNE to the entire dataset, not just to the training set.  The input average SOAP descriptors, $\overline{B}^{(2)}$, clearly exhibit a high degree of structure, to the extent that t-SNE surprisingly reproduces the ternary phase diagram of the alloy.  The first layer of the network reorganizes the data in a hierarchical fashion, grouping structures with the same space group into local clusters.  Within each cluster, the separation of structures according to alloy composition is also maintained.  As a byproduct, we also see that a significant number of high energy structures in the data occur for space groups I\textit{a}$\overline{3}$ and \textit{Fd}$\overline{3}$\textit{m} which contain a significant amount of Al.  At the final hidden layer, this hierarchical structure has been replaced by a single quasi-1D cluster, with an orientation determined by the mixing energy.  This conforms with the intuition provided by a renormalization group interpretation of the network \cite{Vidal:prl.2007, Evenbly:prl.2014, Mehta:arxiv.2014, Lin:jsp.2016}, in the sense that lower levels in the network resolve finer details in the structure representations, and these are subsequently coarse grained in mapping to the final target quantities.

%----------------------------------------------------------------------------------------
\subsection{Compression of large descriptors} \label{tn:compression}
While the expansion (\ref{ACE}) is mathematically well-defined \cite{Bachmayr:arxiv:2019}, the exponential scaling of the descriptors $B^{(\nu)}$ with the number of distinct chemical species and the cutoffs of the radial and angular momentum channels can yield exceptionally large feature spaces.  This can impose computational limitations on the application of machine learning to atomistic modeling, particularly when the structures in the dataset span compositions with a large number of elements.  Thus, recent efforts \cite{Willatt:pccp.2018, Imbalzano:jcp.2018, Artrith:prb.2017, Glielmo:prb.2018, Faber:jcp.2018, Shapeev:cms.2017} have been made toward optimizing the efficiency of atomic representations while maintaining their symmetry invariance.  This was, for example, a central motivation for the introduction of generalized kernel functions \cite{Willatt:jcp.2019}, alchemical or otherwise (cf. Section \ref{tn:other_methods}).

\begin{figure}[!h]
	\centering
		\includegraphics[scale=0.2]{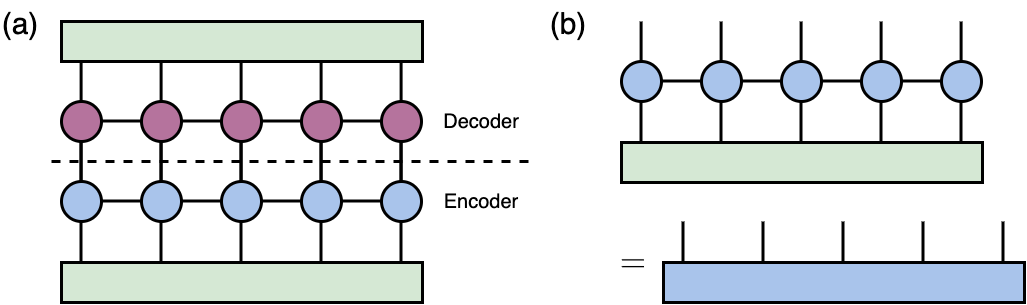}
		\caption{(a) An autoencoder based on MPOs used to compress the physical dimensions of large input descriptor tensors.  (b) A compressed tensor is given by the contraction of the encoder MPO with the original input.}
	\label{autoencoder}
	\end{figure}

As discussed above, one way to alleviate the computational complexity of learning with high dimensional descriptors is to employ TN factorizations of the model architectures, which impose a kind of structured sparsity on the trainable model weights.  Similar constructions can be applied to the input descriptors.  In this section, we explore this possibility with two methods.  First, we validate the MPS factorization scheme (Figure \ref{mps-decomp}) discussed in Section \ref{tensor-facts}.  Second, we implement an autoencoder using MPOs (Figure \ref{autoencoder}).  In principle, these two methods can be combined, both in training the autoencoder and in subsequent supervised learning with the encoded input tensors, although we do not test this combination here.

To test these methods, we use the BA10 dataset \cite{Nyshadham:npj-cm.2019}, which consists of 15950 binary alloys and their mixing energies.  The binary alloys span composition with 10 different metals and 3 different lattice symmetries, face-centered cubic (fcc), body-centered cubic (bcc) and hexagonal close-packed (hcp), with volumes up to 8 atoms per unit cell.  It is worth noting that, like NMD18, the structures in BA10 retain their idealized lattice positions, while the volumes are scaled according to Vegard's law.  However, the DFT mixing energies are computed for these unrelaxed structures, as opposed to the relaxed ground states in NMD18.  Hence, the prediction accuracy in prior studies \cite{Nyshadham:npj-cm.2019, Langer:arxiv.2020} was found to be more limited by the descriptor parameters than implicit noise in the dataset.  In addition to the large number of distinct chemical species, accurate models based on the SOAP power spectrum, $B^{(2)}$, require a fairly large number of radial ($n_\mrm{max} = 8$) and angular momentum ($l_\mrm{max} = 8$) channels \cite{Nyshadham:npj-cm.2019, Langer:arxiv.2020}.  As in \cite{Langer:arxiv.2020}, we will consider training-testing splits of 1600 and 1000 structures, respectively, for both the unsupervised task of training the autoencoder and the subsequent supervised task of predicting the mixing energies.  For the latter supervised task, we use an MPS model with the encoded tensors as input.  Given the exponential scaling of the descriptor dimensions, the input state $| \overline{B}^{(2)} \rangle$ retains only the independent entries of the original descriptor tensor.  Furthermore, zero-padding is added to the state $| \overline{B}^{(2)} \rangle$ to allow for uniform physical dimensions in the subsequent MPS and MPO tensor networks.  Note that this does not affect the symmetry invariance of the descriptor, but the physical dimensions of the MPS and MPO architectures in this analysis do not admit the same correspondence with the structural kernels discussed above.  Nonetheless, the high accuracies obtained with these models underscores the flexibility in choosing the TN factorization and their broader applicability in machine learning \cite{Novikov:arxiv.2015, Stoudenmire:arxiv.2016, Glasser:IEEEa.2020, Gao:prr.2020, Convy:arxiv.2021, Martyn:arxiv.2020, Lu.arxiv.2021}.

\begin{figure}[!t]
	\centering
		\includegraphics[scale=0.35]{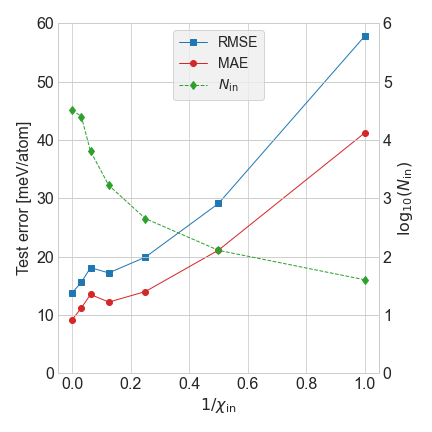}
		\caption{The test performance of a supervised MPS model using MPS factorizations of the dense input descriptors on the BA10 dataset.  The number of input parameters for a MPS factorization with maximum bond dimension $\chi_\mrm{in}$ is also plotted.  The case where the exact input descriptors are used ($1/\chi_\mrm{in} \ra 0$) is included for comparison.}
	\label{mps_decomp_error}
	\end{figure}

Figure \ref{mps_decomp_error} shows the change in the test errors as a function of the maximum bond dimension, $\chi_\mrm{in}$, for a MPS factorization of the descriptor state $| \overline{B}^{(2)} \rangle$.  In this case, the factorized descriptor is used as input to a supervised MPO model ($\chi = 4$).  This is compared to the performance of the unfactorized, dense $\overline{B}^{(2)}$ tensor, and we also plot the corresponding number of input parameters, $N_\mrm{in}$.  As expected, retaining large bond dimensions results in small deviations from the baseline error, while the efficiency of contracting the tensor network is substantially improved.  The test errors grow as the bond dimension decreases, but surprisingly, this growth is nonmonotonic.  Thus, despite reducing the number of input parameters by up to two orders of magnitude, the test errors remain relatively stable, and there are certain smaller bond dimensions that outperform larger ones.  The origin of this nonmonotonic behavior is unclear.  Naively, one would expect that discarding lower singular values at each virtual bond would introduce noise into the inputs, eventually degrading the model performance.  However, it appears that some competing process is present.  One possibility is that compressing the input tensors can improve the quality of the loss landscape (i.e., its smoothness or convexity), similar to overparameterization in deep neural networks \cite{AllenZhu:arxiv.2018, Li:arxiv.2018}.  If that is the case, then deciding how to balance the input factorization against the TN model architecture would be an interesting open question.

We mentioned previously that an MPO applied to an input state can be viewed as a multilinear map from the atom-centered Hilbert space to a potentially lower dimensional space.  Moreover, the topology of certain hierarchical tensor networks, such as TTNs and MERAs, can be viewed as enforcing this kind of dimensional reduction (see Section \ref{tn:latent}).  For instance, it has recently been recognized that there is a fundamental correspondence between MERAs and wavelet transformations \cite{Evenbly:PRL.2016}.  We can exploit this general idea to learn approximately faithful compressions of the input descriptors.  To do so, we employ MPOs in an autoencoder setup, shown in Figure \ref{autoencoder}, where the output dimensions, $d_\mrm{latent}$, of the MPO tensors are less than the input dimensions, $d_\mrm{in}$.  In this setting, the \textit{encoder} MPO maps the input state to a lower dimensional latent space, while the \textit{decoder} MPO attempts to invert this transformation.  The autoencoder network is trained by SGD to minimize the MSE between the components of the input state and its reconstruction by the decoder.  The trained encoder contracted with the input state (Figure \ref{autoencoder}b) yields a compressed tensor adapted to the data.

In Table \ref{ae_errors}, we record the test reconstruction errors (RMSE and MAE) for autoencoders with different $d_\mrm{latent} < d_\mrm{in} = 8$, as well as the test performance of the encoded inputs in predicting the mixing energies with an MPS model.  The bond dimensions of the autoencoder MPOs are fixed ($\chi_\mrm{enc} = \chi_\mrm{dec} = 6$), as are the bond dimensions of the supervised MPS model ($\chi_\mrm{mps} = 10$).  The compression mediated by the encoder is measured against the original size of the input state, $d_\mrm{latent}^n / d_\mrm{in}^n$.  We find that the MPS model performance in predicting the mixing energies remains high even for encoded inputs retaining a few percent of the original number of parameters.  

	\begin{table}
    		\caption{The performance of a 5-site MPO-based autoencoder with virtual bond dimension $\chi_\mrm{enc/dec} = 6$ is measured by the RMSE (MAE) for reconstructing input descriptors in a test set.  This test reconstruction error (r-Error) is shown for varying the latent dimensions, $d_\mrm{latent}$, of the encoder, and the subsequent compression of the input $d_\mrm{latent}^5 / d_\mrm{in}^5$ is also included.  The test RMSE (MAE), denoted s-Error, characterizes the performance of a supervised MPS model ($\chi_\mrm{mps} = 10$) applied to the compressed inputs.}
    		\begin{tabular*}{\columnwidth}{ c  c  c  c  c  c }
     		\hline \hline
      		\ \ $d_\mrm{latent}$ \ \ & \ Compression \ & \ r-Error \ & \ s-Error [meV/atom] \ \\  % & \ \ MAE (test) \\
     	 	\hline
			-  & 1 & - &  14.8 (9.9) \\   
			6 & 0.237 & 0.060 (0.031) & 16.3 (11.2) \\
			4 & 0.031 &  0.088 (0.047) & 19.3 (13.9) \\  
			2 & $9.76 \times 10^{-4}$ & 0.137 (0.070) & 103.2 (77.8) \\
     	 	\hline \hline
    		\end{tabular*}
	\label{ae_errors}
	\end{table}

Thus, we have found that both an MPS decomposition and an MPO-based autoencoder are effective methods to reduce the computational complexity imposed by the input descriptors.  It is again worth emphasizing the parametric efficiency of these methods compared to more standard deep learning architectures.  For example, the deep convolution neural network in the original BA10 work \cite{Nyshadham:npj-cm.2019} uses $\mcl{O}(10^8)$ parameters combined with many-body tensor representations (MBTR) \cite{Huo:arxiv.2017}, which lead to feature vectors with $\mcl{O}(10^5)$ entries.  On the other hand, the high-performing MPS / MPO models possess $\mcl{O}(10^3)$ parameters with compressed input tensors of similar size.

%----------------------------------------------------------------------------------------
%----------------------------------------------------------------------------------------
\section{Discussion and Summary}
In this work, we have demonstrated that tensor networks provide both a unifying formalism for a large class of atomic structure representations as well as robust machine learning architectures based on them.  Specifically, we detailed how the ACE framework admits a natural description in terms of symmetric tensor networks.  This formalism allows for the transparent application of SO(3) recoupling theory to the construction of a rotationally invariant basis and generalizes straightforwardly to SO(3)-equivariant cases.  By taking seriously the Hilbert space structure of atom-centered descriptors \cite{Willatt:jcp.2019}, we have shown how tools commonly applied in quantum many-body physics can be repurposed in coarse-grained surrogate models of fundamental material properties.  

Indeed, the map between a solid-state or molecular structure and a given property can be written as an inner product between a learnable state and a sufficiently representative descriptor, both of which admit low-rank tensor network factorizations.  We have found that, as learnable architectures, MPOs and MPSs exhibit strong generalizability in common learning tasks in material informatics, and their performance is particularly notable when the amount of available data is very limited.  We provided two routes toward understanding this apparent inductive bias.  First, we showed how learnable tensor networks, viewed as multilinear maps on the atom-centered Hilbert space, can be related to generalized kernel functions.  An appropriate choice of network topology thus provides an explicit way of coupling structural and chemical components in the model.  Unlike standard structural kernel methods, the evaluation of these TN models scales more efficiently with the number of structures.  This is similar to the application of neural networks, and the MPS models benchmarked in this work can be viewed as parametrically efficient regularizations of a dense neural network.  Indeed, the apparent compressibility of the descriptor space and the learning architecture can be used to obtain more efficient atomic representations, and we demonstrated this using both a standard MPS decomposition algorithm as well as a MPO-based autoencoder.  The second way to understand the strong generalizability of these TN models is more fundamental and potentially a universal feature of tensor-network machine learning more broadly.  In this case, the learned TN states possessed signatures of high entanglement complexity.  In particular, the initial WD statistics of the level spacings in the entanglement spectrum were essentially preserved under stochastic gradient descent, while the spectrum itself adapted to the learning task.

The above discussion points to several directions for potential future work.  First, general TN factorizations can be used to replace fully connected layers common in deep learning architectures, a program already undertaken in \cite{Novikov:arxiv.2015, Gao:prr.2020}.  In the context of materials prediction, adding TN layers in end-to-end architectures like graph neural networks is an effective way to reduce memory overhead and could potentially improve performance on smaller training sets.  Similarly, TN layers could be implemented in the Behler-Parrinello neural network framework \cite{Behler:prl.2007, Behler:chemrev.2021} for modeling high-dimensional potential energy surfaces.  In this case, separate tensor networks would replace the neural networks applied to each local environment.  On a related note, it would be worth quantifying the performance of these networks on very large datasets and identifying, in particular, whether there is some advantage to using deep versus shallow network architectures \cite{Lin:jsp.2016}.

Let us mention that training TNs with a large number of tensors, as would be the case (\ref{copy_ops}) for high orders $\nu$ and $\zeta$, using SGD suffers from the presence of barren plateaus in the loss function \cite{Liu:arxiv.2021, McClean:NatComm.2018, Cerezo:NatComm.2021}.  For a large MPS, this is intrinsically related to the fact that, under random initialization, the state constitutes an approximate 2-design for which the expected gradients vanish \cite{Liu:arxiv.2021}.  It may be possible to overcome this problem by choosing a different initialization scheme \cite{Haferkamp:PRXQ.2021}, or by preconditioning the network with a different algorithm, such as DMRG.  Formulating a local loss function should also alleviate the issue \cite{Liu:arxiv.2021, Cerezo:NatComm.2021}.  Alternatively, one could utilize common strategies employed in deep learning for regulating gradients.  For example, a local tensor network scanned across sub-partitions of an input tensor provides a TN analogue of weight sharing in a convolutional neural network.  Residual skip connections and batch normalization may also prove valuable in deeper networks.  A related question underlying these approaches is the extent to which they lead to complexly entangled states.  If WD statistics in the entanglement spectrum is a fundamental signature of a highly generalizable model, it would be worth characterizing, in general, how the entanglement spectrum in approximate $k$-designs evolves under stochastic gradient descent.

From a practical standpoint, the parametric efficiency of these methods as well as their high performance on limited training data makes a strong case for their application in large-scale materials simulation and high-throughput screening, particularly when the calculation of target parameters from first principles is computationally demanding.  Furthermore, while we have characterized these methods on the important class of atom-density representations, we expect that they are broadly adaptable to generic material feature spaces \cite{Onat:jcp.2020}.

%----------------------------------------------------------------------------------------
%----------------------------------------------------------------------------------------
\appendix

%----------------------------------------------------------------------------------------
\section{Tensor network graphical notation} \label{appendix:tn-basics}

\begin{figure}[!h]
	\centering
		\includegraphics[scale=0.25]{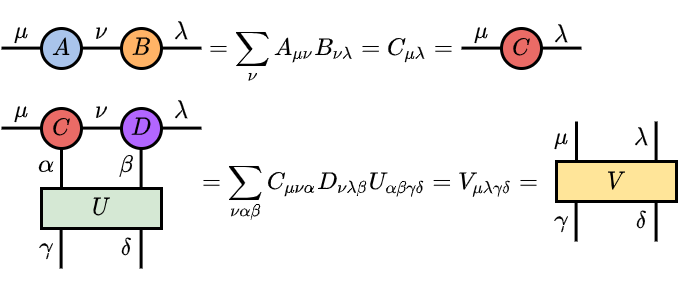}
		\caption{Examples of graphical notation for tensor networks.  Tensor indices associated with each tensor in the network are represented by open edges.  Closed edges between pairs of tensors indicate summation / contraction over the matching tensor indices.}
	\label{tn-notation}
	\end{figure}

An arbitrary state $|\psi \rangle = \sum_{\{s\}} \psi_{s_1 s_2\dotsm s_n} | s_1 s_2 \dotsm s_n \rangle$ in a tensor product space $| \psi \rangle \in \mcl{H}^{\otimes n}$, where each $| s_k \rangle$ forms a basis for a finite-dimensional vector space $\mcl{H}$, can be described by the tensor $\psi_{s_1 s_2\dotsm s_n}$ formed by its coefficients.  In the standard diagrammatic notation of tensor networks, an arbitrary, dense tensor, $\psi_{s_1 s_2\dotsm s_n}$, of order $n$ is represented by a shape or node with $n$ open edges.  Summation over a matching pair of tensor indices, otherwise referred to as \textit{contraction}, is represented by a closed edge in the network.  The familiar example of matrix multiplication is shown as a tensor network in Figure \ref{tn-notation}, as well as a less trivial network.

%----------------------------------------------------------------------------------------
\section{Additional SO(3) recoupling relations} \label{appendix:recoupling}

\begin{figure}[!h]
	\centering
		\includegraphics[scale=0.25]{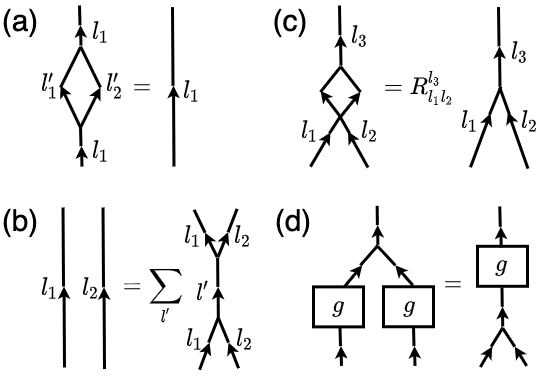}
		\caption{Additional recoupling relationships between Clebsch-Gordan tensors and Wigner $D$-matrices, including (a) orthogonality, (b) completeness, and (d) naturality /equivariance.  (c) Braiding between irrep edges is defined in terms of a set of $R$-symbols, which reduce to sign factors in SO(3).}
	\label{CG_identities}
	\end{figure}

In this appendix, we collect some useful identities in the algebra of SO(3) representations.  Recalling the diagrammatics from the main text, each line with an arrow is labelled by an irrep $l$ and spans an associated subspace of dimension $d_l = 2l + 1$.  Intertwiners between irreps are given by the usual Clebsch-Gordan (CG) coefficients (Figure \ref{basis_network_comps}c,d), which satisfy the orthogonality relations shown in Figure \ref{CG_identities}a,b.  The CG tensors commute with the action of the group (Figure \ref{CG_identities}d for $C$ and analogous for $C^\dagger$); that is, they form a natural (i.e., equivariant) transformation.  Note that the following three identities are sufficient to derive the generalized Wigner-Eckhart theorem discussed in the main text (Figure \ref{JLV_symmetry}).

\begin{figure}[!h]
	\centering
		\includegraphics[scale=0.25]{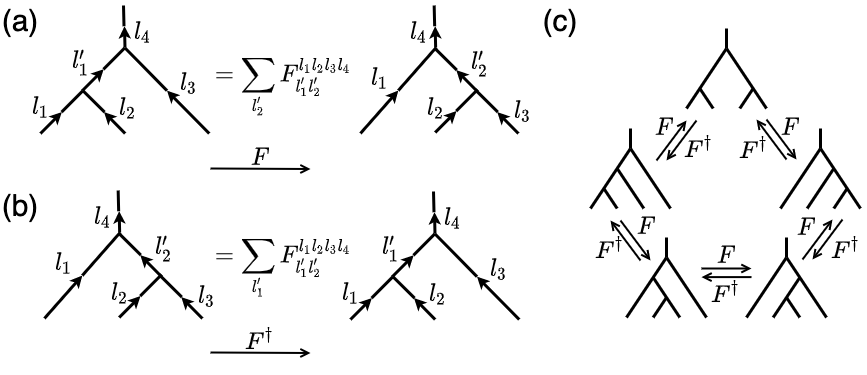}
		\caption{Unitary transformations (a,b) known as $F$-symbols define mappings between recoupling schemes.  They must satisfy the consistency condition (c) known as the pentagon identity.}
	\label{F-symbols}
	\end{figure}

The recoupling of three or more irreps can be described by the contraction of multiple CG tensors to form a fusion tree, and recoupling schemes represented by different fusion trees can be related to each other via unitary transformations known as $F$-symbols (Figure \ref{F-symbols}a,b).  In SO(3), explicit values for the $F$-symbols can be derived from the Wigner $6j$ symbols (or alternatively from Racah $W$ coefficients), which are given by the maximally irreducible diagram formed by the contraction of two 4-valent fusion trees \cite{Stedman:dtgt, Cvitanovic:gt}.  Additionally, pairs of irreps can be swapped  (Figure \ref{CG_identities}c) using the so-called $R$-symbols.  In SO(3), they take the explicit values $R_{l_1 l_2}^{l_3} = (-1)^{l_1+ l_2 - l_3}$ and can be used to constrain to the parity of the state.  For instance, an SO(3)-invariant $\nu$-order descriptor possesses inversion symmetry if $\sum_{k=1}^\nu l_k$ is even, which corresponds to $\left(\prod_{k=1}^{\nu-2} R_{l'_k l_{k+2}}^{l'_{k+1}} \right) R_{l_1 l_2}^{l'_1} = 1$ for $\nu>2$ and $l'_{\nu-1}=0$ in the recoupling scheme used in the main text.

\begin{figure}[!t]
	\centering
		\includegraphics[scale=0.25]{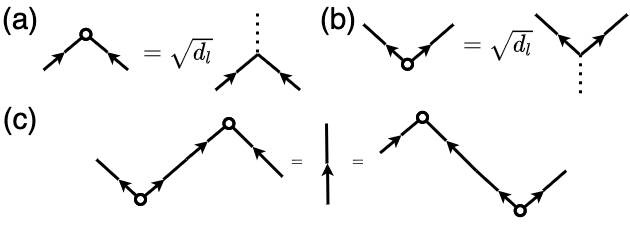}
		\caption{(a) Cap and (b) cup tensors defined in terms of Wigner $2jm$ symbols.  Diagrammatically, they allow for the reversal of an irrep edge orientation, and satisfy the inversion identity (c).}
	\label{cups-caps}
	\end{figure}

Let us mention that additional coherence relations must be satisfied by the $F$- and $R$-symbols, namely the so-called pentagon and hexagon identities \cite{Wang:topological.2010}, where the pentagon identity is shown in Figure \ref{F-symbols}c.  Indeed, the algebraic structure of finite-dimensional representations of SO(3) corresponds more generally to a symmetric tensor category.  A useful aspect of this construction is the ability to encode nontrivial manipulations of tensors in the local deformations of diagrams.  For example, the reversal an irrep arrow can be accomplished by contracting with normalized Wigner $2jm$ symbols, $\epsilon^{(l)}_{mm'} \coloneqq \sqrt{d_l} C^{(l l 0)}_{m m' 0} = (-1)^{l-m} \delta_{m,-m'}$, where the \textit{cup} and \textit{cap} tensors $\epsilon^{(l)}$, $(\epsilon^{(l)})^\dagger$ are matrix inverses of each other, $\epsilon^{(l)} (\epsilon^{(l)})^\dagger = (\epsilon^{(l)})^\dagger \epsilon^{(l)} = \mathbb{I}^{(l)}$ (Figure \ref{cups-caps}).  The cup and cap tensors will be used to recursively construct equivariant descriptors in the following section.
	
%----------------------------------------------------------------------------------------	
\section{Generalizing to SO(3) equivariants and beyond} \label{appendix:equivariants}
In this appendix, we show how the SO(3)-invariant tensor networks in Section \ref{descriptor-calc} can be generalized to SO(3)-equivariance and applied to the learning of vectorial and tensorial properties.  In doing so, we will largely reproduce central results related to the general ACE formalism \cite{Drautz:prb.2020, Bachmayr:arxiv:2019}, as well as the so-called $\lambda$-SOAP \cite{Grisafi:prl.2017} descriptors and the recursive N-body iterative contraction of equivariants (NICE) framework \cite{Nigam:jcp.2020}.

First, we recall from the main text (Figure \ref{JLV_symmetry}) that $\nu$-order SO(3)-invariant tensors can be explicitly constructed by taking the Haar integral over the action of the group.  This can be understood as a projection onto the invariant subspace, $\mrm{Inv}(\mcl{V}_{l_1} \otimes \dotsb \otimes \mcl{V}_{l_\nu} ) \cong \mrm{Hom}(\mcl{V}_{l_1} \otimes \dotsb \otimes \mcl{V}_{l_\nu}, \mbf{1} )$, of the tensor product of $\nu$ irrep spaces $\mcl{V}_{l_k}$.  This projection operator,

	\begin{align}
		P^{(l_1 \cdots l_\nu)} &= \int dg D^{(l_1)}_{m'_1 m_1}(g) \dotsm D^{(l_\nu)}_{m'_\nu m_\nu}(g) \\
		&= \sum_{\iota} \overline{\iota_{m'_1 \dotsm m'_\nu}} \iota_{m_1 \dotsm m_\nu} \label{inv_decomp}
	\end{align}
	
\begin{figure}[!h]
	\centering
		\includegraphics[scale=0.25]{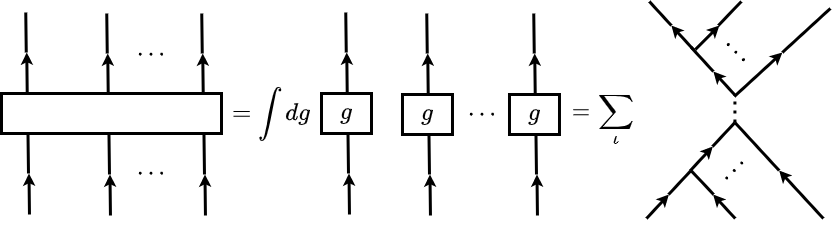}
		\caption{General projection operator onto the invariant subspace $\mrm{Inv}(\mcl{V}_{l_1} \otimes \dotsb \otimes \mcl{V}_{l_\nu} )$ of an arbitrary tensor product space $\mcl{V}_{l_1} \otimes \dotsb \otimes \mcl{V}_{l_\nu}$.}
	\label{haar_proj}
	\end{figure}

\nn
is shown graphically in Figure \ref{haar_proj}, where for $\mcl{G}$ a semisimple group this projection decomposes, as before, into dual fusion trees (i.e., higher-order intertwiners) factored through the trivial irrep space \cite{Oeckl:dgt, Perez:lrr.2013}.  Note that this decomposition is a direct consequence of the Peter-Weyl theorem \cite{Oeckl:dgt}.  We denote these general intertwiners by $\iota_{m_1 \dotsm m_n}$, and the sum in (\ref{inv_decomp}) runs over the intermediate irreps, which depend on the chosen recoupling scheme.

\begin{figure}[!h]
	\centering
		\includegraphics[scale=0.25]{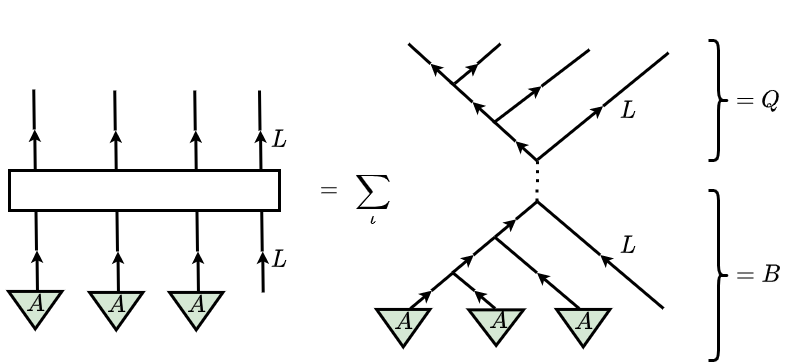}
		\caption{Elements of the basic recursive construction of order-($\nu+1$) equivariant descriptors, involving (a) the reorientation of the final momentum irrep edges using cup tensors and (b) the fusion of additional order-1 descriptors with an order-$\nu$ descriptor.}
	\label{covariant_construct}
	\end{figure}

Under this projection, a $\nu$-order SO(3)-equivariant descriptor, $B^{(\nu)}_L$, transforming as a state with definite angular momentum $L$ can be constructed by symmetrizing over the product of $\nu$ atom-centered tensors, $A_{\alpha n l m}$, along with an identity operator carrying the target irrep space $L$, as shown in Figure \ref{covariant_construct}.  As before, the descriptor, $B^{(\nu)}_L$, corresponds to the reduced part of the decomposition.  By bending upward the open $L$ edge of the corresponding tensor network using cup and cap tensors, one can verify that the tensor product of $\nu$ irreps is mapped to the target $L$ irrep, and the fully invariant case from the main text can be recovered by taking $L=0$.

\begin{figure}[!h]
	\centering
		\includegraphics[scale=0.25]{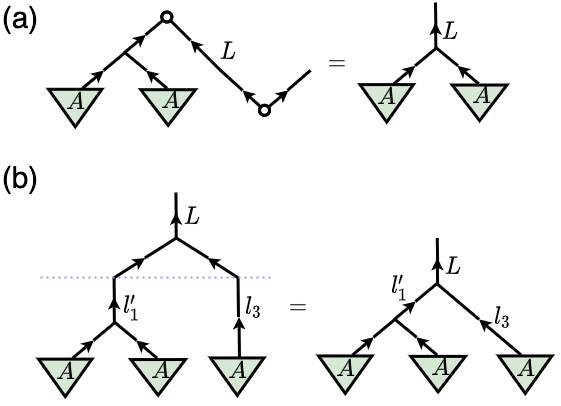}
		\caption{Elements of the basic recursive construction of order-($\nu+1$) equivariant descriptors, involving (a) the reorientation of the final momentum irrep edges using cup tensors and (b) the fusion of additional order-1 descriptors with an order-$\nu$ descriptor.}
	\label{equiv_recursion}
	\end{figure}

From this picture, it is clear that a ($\nu+1$)-order descriptor can be obtained by fusing together lower order descriptors.  This is shown in our chosen recoupling scheme in Figure \ref{equiv_recursion}, where an order-1 descriptor, equivalent to the unsymmetrized atom-centered tensor $A$, is combined with an order-2 descriptor by contraction with the necessary CG, cup and cap tensors.  Repeated application of this process yields a recursive formula for higher order descriptors.

\begin{figure}[!h]
	\centering
		\includegraphics[scale=0.2]{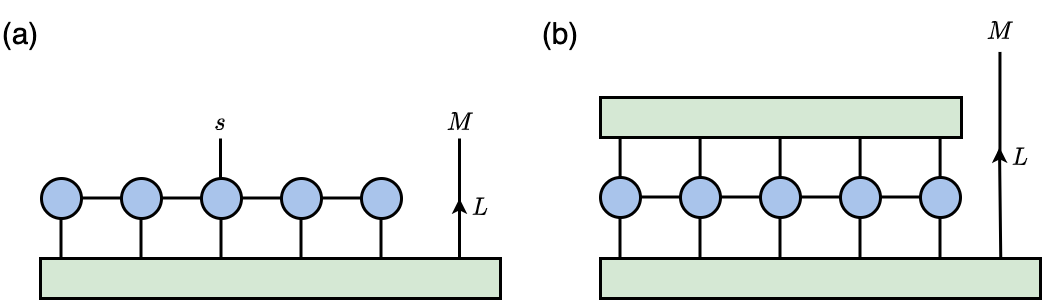}
		\caption{Tensor network models which preserve the equivariance of the input descriptor.  (a) An example MPS model which allows for additional output degrees of freedom $s$.  (b) An example MPO model which implicitly increases the effective many-body character of the input descriptor by contracting with an equivariant descriptor and its invariant counterpart.}
	\label{covariant_model}
	\end{figure}

The angular momentum $L$ of an equivariant descriptor can be preserved by a tensor network model by acting only on the remaining $\{\alpha_k, n_k, l_k\}$ degrees of freedom.  For example, an MPS-like model which maps to a target quantity $y^{(L)}_{sM}$ living in the space $| s \rangle \otimes | LM \rangle$, where $s$ labels a possible additional degree of freedom, is shown in Figure \ref{covariant_model}a.  Naively taking $\zeta$-order tensor products of $L > 0$ equivariant descriptors destroys the angular momentum channel of the original state since products of irreps can be subsequently recoupled.  Alternatively, as mentioned in \cite{Willatt:jcp.2019}, the effective many-body character of an equivariant model can be increased by taking tensor products of the equivariant state, $B^{(\nu)}_L$, with $\zeta-1$ copies its invariant counterpart, $B^{(\nu)}_{L=0}$.  An example $\zeta=2$ MPO-type model is shown in Figure \ref{covariant_model}b.

A useful aspect of this formalism is that the fusion trees that encode the equivariance of the model can be manipulated purely algebraically and computed independently from the atom-centered $A$-tensors \cite{Schmoll:ap.2020}.  As demonstrated in \cite{Bachmayr:arxiv:2019, Lysogorskiy:npj_cm.2021} with the ACE formalism, this allows for potential speedups in applications, such as molecular dynamics or Monte Carlo, which constantly update the underlying descriptors.  In particular, the fusion trees can be recursively precomputed and contracted with the $A$-tensors along with the learned model weights at runtime.

\begin{figure}[!h]
	\centering
		\includegraphics[scale=0.2]{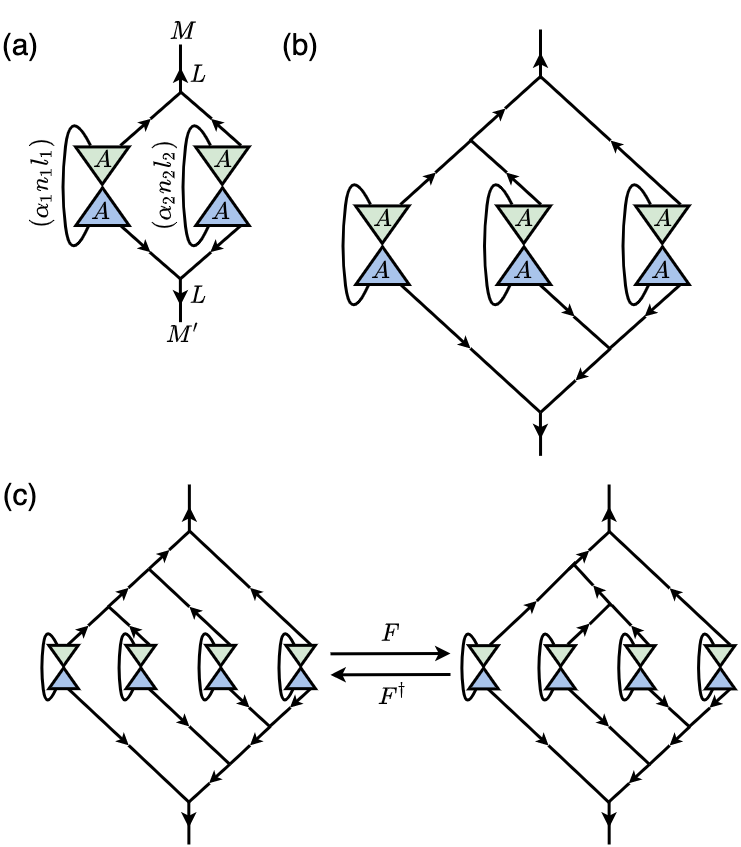}
		\caption{Equivariant $\lambda$-SOAP generalization of the (a) power spectrum ($\nu=2$) and (b) bispectrum ($\nu=3$) structural kernels as tensor networks.  (c) Topologically distinct kernel functions exists for either ($L=0$, $\nu>3$) or ($L>0$, $\nu>2$), determined by the repeated application of $F$-symbols to the pairs of descriptor fusion trees.}
	\label{covariant_kernels}
	\end{figure}

Let us point out some implications for the constructions of equivariant structural kernels \cite{Grisafi:prl.2017}.  Figure \ref{covariant_kernels}a,b displays the diagrams for kernel functions corresponding to the $\lambda$-SOAP generalization of the power spectrum ($\nu=2$) and bispectrum ($\nu=3$), respectively.  The pairs of descriptors in these kernels are built from the same recoupling scheme, and for $L=0$ it can be seen that each kernel function is essentially given by a spin network equivalent to a Wigner $3nj$ symbol  \cite{Cvitanovic:gt} weighted by the trace over the remaining $\{\alpha_k, n_k, l_k\}$ degrees of freedom.  For either ($L=0$, $\nu>3$) or ($L>0$, $\nu>2$), there exist topologically distinct kernel functions determined by the application of $F$-symbols to the pairs of descriptor fusion trees.  An example for $\nu=4$, valid for all $L$, is shown in Figure \ref{covariant_kernels}c, and for $L=0$ the right-hand side is equivalent to the maximally irreducible diagram of a weighted Wigner $6j$ symbol.  This raises the interesting question of whether certain nonequivalent inner products for higher $\nu$ equivariants play a more privileged role than others, particularly as a basis for machine learning tasks.

\begin{figure}[!b]
	\centering
		\includegraphics[scale=0.25]{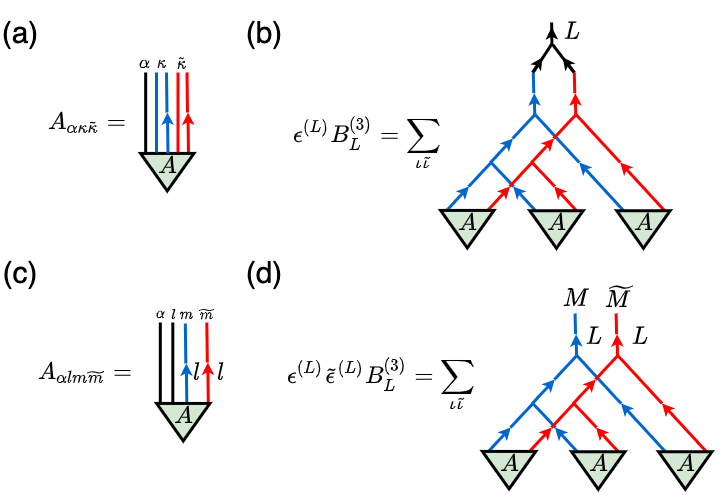}
		\caption{Recoupling schemes for (a-b) SO(3) $\otimes$ SO(3) and (c-d) SO(4) equivariant descriptors.}
	\label{general_covariants}
	\end{figure}

The above discussion has focused on the physically important case of global SO(3) covariance.  However, this recoupling structure can be further generalized.  For example, as described in \cite{Drautz:prb.2020}, the inclusion of classical on-site magnetic moments, $\mbf{m}_i$, either through spin or orbital degrees of freedom can be accommodated by extending the symmetry of the atom-centered descriptors to SO(3) $\otimes$ SO(3).  Note that while bond, $\kappa = (n l m)$, and magnetic, $\widetilde{\kappa} = (\widetilde{n} \widetilde{l} \widetilde{m})$, degrees of freedom transform independently in the atom-centered tensors, $A_{\alpha \kappa \widetilde{\kappa}}$, they must be recoupled to form a descriptor with definite, total angular momentum.  The general construction of such a combined equivariant descriptor is shown in Figure \ref{general_covariants}a,b.  Another example is provided by an extension of the SOAP method to the SO(4) group \cite{Bartok:prb.2013}, otherwise known as the SNAP method \cite{Thompson:jcp.2015}, which circumvents the need to explicitly specify a radial basis function, $R_{nl}(r)$, by embedding the radial degree of freedom into the 3-sphere, $S^3$.  This amounts to treating the radial component as an additional polar angle that, along with the two angles inherited from $S^2$, describes rotations in $\mathbb{R}^4$.  The hyperspherical harmonic functions, $U^{l}_{m\widetilde{m}}$, form a complete Fourier basis for functions on $S^3$.  Since, at the level of Lie algebras, there exists an isomorphism SO(4) $\sim$ SO(3) $\otimes$ SO(3), the SO(4) recoupling CG coefficients, $H$, decompose into products of SO(3) CG coefficients, $H^{(l_1 l_2 l_3)}_{m_1 \widetilde{m}_1 m_2 \widetilde{m}_2 m_3 \widetilde{m}_3} = C^{(l_1 l_2 l_3)}_{m_1  m_2 m_3} C^{(l_1 l_2 l_3)}_{\widetilde{m}_1 \widetilde{m}_2 \widetilde{m}_3}$ \cite{Meremianin:jpa-mg.2006, Caprio:jmp.2010, Bartok:prb.2013}.  Recoupling in the SNAP bispectrum using this "parabolic-type" decomposition \cite{Meremianin:jpa-mg.2006} is shown in Figure \ref{general_covariants}c,d.

	\begin{figure}[!b]
	\centering
		\includegraphics[scale=0.15]{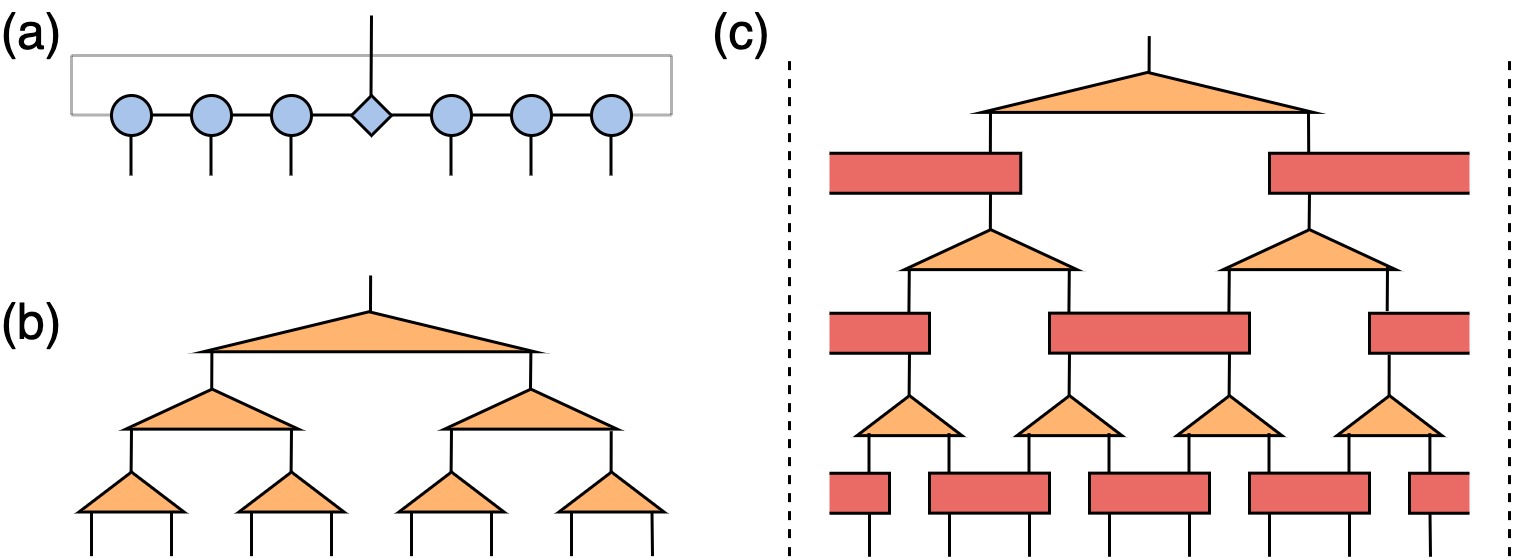}
		\caption{Tensor network factorizations corresponding to (a) a matrix product state (MPS), (b) a tree tensor network (TTN), and (c) a Multi-Scale Entanglement Renormalization Ansatz (MERA).}
	\label{various_tns}
	\end{figure}

%----------------------------------------------------------------------------------------
\section{Additional computational details}  \label{appendix:comp_details}
SOAP power spectra were computed with the \textsc{Dscribe} software package \cite{Himanen:cpc.2020}.  GPR models were trained with the \textsc{Scikit-Learn} library \cite{scikit-learn} using expectation maximization.  The construction and training of deep neural networks and tensor network models were facilitated by the \textsc{TensorNetwork} \cite{tensornetwork2019} and \textsc{TensorFlow} \cite{tensorflow2015} libraries.  These models were trained using the Adam optimization algorithm, a variant of stochastic gradient descent that adaptively updates the learning rates based on moments of the gradients \cite{Kingma:arxiv.2014}.

\begin{acknowledgments}
This research was supported  through the UW Molecular Engineering Materials Center, a Materials Research Science and Engineering Center (Grant No. DMR-1719797) and was facilitated through the use of advanced computational, storage, and networking infrastructure provided by the Hyak supercomputer system and funded by the STF at the University of Washington. \end{acknowledgments}

%\bibliography{ml-tn_refs}

\end{document}